%% file: main.tex
\documentclass[acmlarge,nonacm]{acmart}

\usepackage{graphicx}
\usepackage{tikz}
\usepackage{xspace}
\usepackage{comment}
\usepackage{multirow}
\usepackage{subcaption}

\newcommand{\name}{TRAMBA\xspace}
\newcommand{\todo}[1]{\ClassWarning{NOT READY TO SUBMIT}{There is something left todo} \textcolor{red}{[#1]}}

\AtBeginDocument{%
  \providecommand\BibTeX{{%
    \normalfont B\kern-0.5em{\scshape i\kern-0.25em b}\kern-0.8em\TeX}}}

\setcopyright{acmlicensed}
\copyrightyear{2018}
\acmYear{2018}
\acmDOI{XXXXXXX.XXXXXXX}

\acmJournal{IMWUT}
\acmVolume{37}
\acmNumber{4}
\acmArticle{111}
\acmMonth{8}

\begin{document}

\title[\name: A Hybrid Transformer and Mamba Architecture for Acoustic Speech Enhancement]{\name: A Hybrid Transformer and Mamba Architecture for Practical Audio and Bone Conduction Speech Super Resolution and Enhancement on Mobile and Wearable Platforms}

\author{Yueyuan Sui}
\authornote{Both authors contributed equally to this work.}
\affiliation{%
  \institution{Northwestern University}
  \streetaddress{633 Clark Street}
  \city{Evanston}
  \state{Illinois}
  \country{United States}}
\email{yueyuansui2024@u.northwestern.edu}

\author{Minghui Zhao}
\authornotemark[1]
\affiliation{%
  \institution{Columbia University}
  \streetaddress{116th and Broadway}
  \city{New York}
  \state{New York}
  \country{United States}}
\email{mz2866@columbia.edu}

\author{Junxi Xia}
\affiliation{%
  \institution{Northwestern University}
  \streetaddress{633 Clark Street}
  \city{Evanston}
  \state{Illinois}
  \country{United States}}
\email{junxixia2024@u.northwestern.edu}

\author{Xiaofan Jiang}
\affiliation{%
  \institution{Columbia University}
  \streetaddress{116th and Broadway}
  \city{New York}
  \state{New York}
  \country{United States}}
\email{jiang@ee.columbia.edu}

\author{Stephen Xia}
\affiliation{%
  \institution{Northwestern University}
  \streetaddress{633 Clark Street}
  \city{Evanston}
  \state{Illinois}
  \country{United States}}
\email{stephen.xia@northwestern.edu}

\renewcommand{\shortauthors}{Sui et al.}

\begin{abstract}
We propose \name, a hybrid transformer and Mamba architecture for acoustic and bone conduction speech enhancement, suitable for mobile and wearable platforms. Bone conduction speech enhancement has been impractical to adopt in mobile and wearable platforms for several reasons: (i) data collection is labor-intensive, resulting in scarcity; (ii) there exists a performance gap between state-of-art models with memory footprints of hundreds of MBs and methods better suited for resource-constrained systems. To adapt \name to vibration-based sensing modalities, we pre-train \name with audio speech datasets that are widely available. Then, users fine-tune with a small amount of bone conduction data. \name outperforms state-of-art GANs by up to 7.3\% in PESQ and 1.8\% in STOI, with an order of magnitude smaller memory footprint and an inference speed up of up to $465$ times. We integrate \name into real systems and show that \name (i) improves battery life of wearables by up to 160\% by requiring less data sampling and transmission; (ii) generates higher quality voice in noisy environments than over-the-air speech; (iii) requires a memory footprint of less than 20.0 MB.
\end{abstract}

\ccsdesc[500]{Computer systems organization~Embedded and cyber-physical systems}
\ccsdesc[500]{Human-centered computing~Ubiquitous and mobile computing}
\ccsdesc[500]{Computing methodologies~Artificial intelligence}
\ccsdesc[500]{Computing methodologies~Machine learning}

\keywords{Audio Super Resolution, Speech Enhancement, Bone Conduction Enhancement, Transformer, Mamba, Earable Computing}


\maketitle

\input{sections/introduction}

\input{sections/related_works}

\input{sections/superresolution}
\input{sections/nn_evaluation}

\input{sections/system_design}
\input{sections/end_to_end_evaluation}

\input{sections/discussion}
\input{sections/conclusion}


\bibliographystyle{ACM-Reference-Format}
\bibliography{sample-base}



\end{document}

%% file: sections/introduction.tex
\section{Introduction}
\label{sec:introduction}


\begin{figure}[!t]
    \centering
    \includegraphics[width=0.50\linewidth]{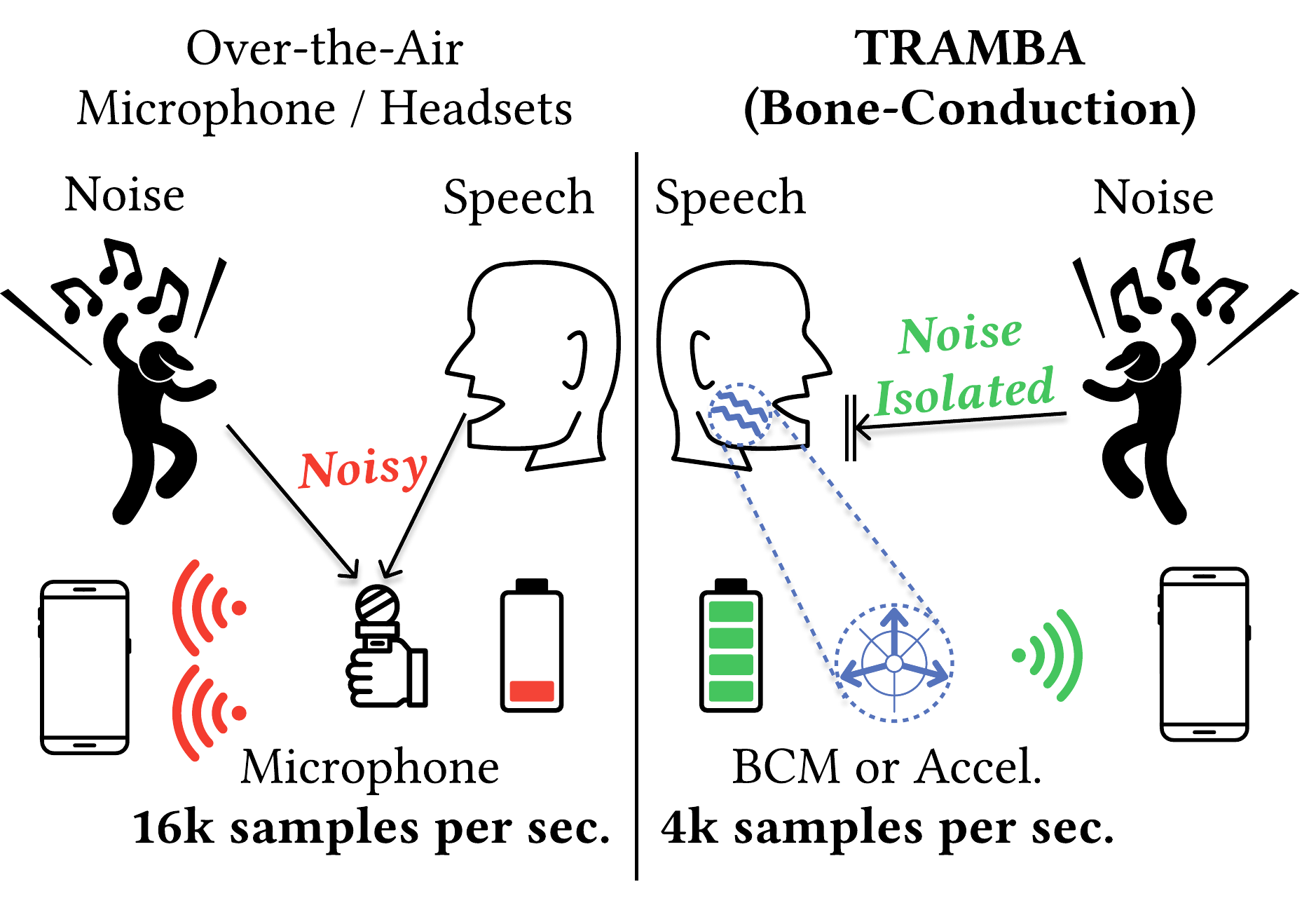}
    \caption{\name enhances vibration-based speech, which is naturally insensitive to ambient noises.}
    \label{fig:application}
\end{figure}

Wearables have revolutionized the way we interact with technology and have enabled continuous and real-time monitoring of our health and wellness. Since the early 2000s, wearables have continued to play an increasingly important role in our lives. The wearables market is expected to grow from $70$ billion USD in 2023 to $230$ billion USD by 2032~\cite{market2024wearable}. Head-worn wearables, including earphones and glasses, continue to be one of the fastest growing segments ($71$ billion USD in 2023 to $172$ billion USD by 2030~\cite{gvr2024earphones,gvr2023glasses}). This has been spurred on, in part, by the increasing importance and prevalence of earables, augmented reality (AR), and virtual reality (VR).

One unique signal that can only be sensed by wearables near or directly on the head is speech. Traditionally, speech is captured by over-the-air (OTA) microphones that convert variations in air pressure into an electrical signal, which can be used for a variety of important applications (e.g., phone conversations and voice commands). OTA microphones are commonly found on earbuds or headphones and are generally positioned near the user’s mouth. However, OTA microphones easily pick up background sounds, especially in noisier environments, which can degrade speech quality.

To solve this problem, there have been many works that explore denoising, sound source separation, and speech enhancement methods to separate speech from background noise~\cite{vincent2018audio,nugraha2016multichannel,luo2018tasnet,luo2019conv,rethage2018wavenet,huang2022investigating}. However, this approach is challenging because 1) the model is ``blind’’ to the types of background noises that may occur. There could be any number of different sounds, or even another person’s speech, that the model needs to account for without prior knowledge, while preserving the wearer’s speech. 2) It is not uncommon to be in areas where background noise can easily interfere and overpower the user’s speech (e.g., a busy cafeteria or construction), which makes speech extraction and enhancement extremely difficult.

Unlike OTA microphones, bone conduction microphones (BCM) are placed in direct contact with the head. BCMs are sensitive to vibrations produced by the skin and skull as the person speaks and are not sensitive to changes in air pressure, making them naturally robust against ambient noise. Other sensors that can measure vibration and forces of motion, such as accelerometers (ACCEL), have shown sensing capabilities of speech at lower fidelities~\cite{he2023towards,liang2022accmyrinx} and of general facial movements for other important applications, such as authentication~\cite{srivastava2022muteit}. Bone conduction sensing methods are naturally not sensitive to ambient background noises compared to OTA microphones. However, vibration-based acoustic modalities see \textit{severe attenuation at higher frequencies, which significantly degrades intelligibility and speech quality}. 

There are several works that explore vibration and bone-conduction super resolution methods to reconstruct higher frequencies and improve speech quality~\cite{he2023towards,li2022enabling}, but do not consider practical aspects for real-time mobile, wearable, and earable systems, as detailed below:
\begin{enumerate}
\item \textbf{Heavy Processing:} State-of-art speech super resolution models, namely generative adversarial networks (GANs), require at least tens of millions of parameters (several hundred MB). This makes loading, running, and finetuning the model less practical. 
\item \textbf{Performance Gap:} Methods with smaller memory footprints that target mobile platforms with less computational resources (e.g., many U-Net architectures~\cite{li2022enabling,liu2018bone}) see a large gap in performance compared to top performing models, which \name addresses.
\item \textbf{Lack of data:} Training a super resolution model for vibration-based sensors requires paired vibration (BCM or ACCEL) and OTA microphone data that captures the higher frequencies. There is a dearth of this data publicly available. Additionally, collecting this data is very labor intensive because it requires building a contraption that records both BCM/ACCEL data synced with a microphone that captures the higher frequencies. (i) Most super resolution works focus on OTA microphones, where low-fidelity signals can be generated simply by low pass filtering clean speech, for which there are many datasets for. Applying a model trained on such a dataset performs poorly on speech captured by vibration-based sensors. (ii) Other works that focus on bone-conduction settings make commendable efforts to painstakingly collect data from a limited set of volunteers or artificially generate data from clean speech audio. Training a model with limited real samples does not generalize well. We demonstrate both of these shortcomings in Section~\ref{sec:eval_transfer_learning}.
\item \textbf{System Optimization:} Considerations such as sampling rate and where to perform computation can have significant impact on inference time and battery life of the wearable.
\end{enumerate}

Leveraging recent advances in attention and state space models, we propose \name, a hybrid transformer and Mamba-based model for acoustic speech enhancement and super resolution. \name outperforms state-of-art GANs by up to 7.3\% in PESQ and 1.8\% in STOI, with an order of magnitude smaller memory footprint and an inference speed up of up to $465$ times. Compared to state-of-art methods, namely GANs, \name only requires 5.2 million parameters, compared to a minimum of tens of billions of parameters. This yields a model size in the order of MBs compared to hundreds of MBs or more when implemented on a mobile phone. Figure~\ref{fig:application} highlights the advantages of \name compared to OTA methods.

To adapt \name to bone conduction sensing modalities, we first pre-train \name for super resolution on speech collected from OTA microphones, before fine-tuning the model with a small amount of data collected from the user. Building the model in this way allows us to take advantage of the vast amount of standard audio datasets that are publicly available, instead of relying on a large amount of paired audio and vibration data, which is both scarcely available and labor intensive to collect. In real systems, this data collection can occur only one time upon box opening. We show that \name outperforms state-of-art speech enhancement GANs on BCMs and accelerometers across different users and placements on the face by up to 5.9\% for PESQ, 1.8\% for STOI, while requiring almost 20 times less time to fine-tune on only 15 minutes of data collected from an individual. The amount of data required is lower, if not on-par, with other voice generation applications that need to collect voice samples from users, such as~\cite{create2023apple} from Apple.

We incorporate \name into an end-to-end wearable and mobile platform and demonstrate, through user studies, that \name improves word error rate by up to $75$\% in noisy environments compared to approaches that aim to suppress noise in OTA speech. Additionally, we demonstrate that the lack of high-frequency components in bone conduction-based sensing modalities allows us to significantly reduce both the sampling rate of the sensor and the transmission rate, which can improve the battery life of wearables by up to $160$\%.

This work introduces a novel speech enhancement method that outperforms existing state-of-art methods, with a memory footprint that is orders of magnitude smaller. To the best of our knowledge, this is also the first work that reconstructs intelligible speech using only a single wearable accelerometer. Our contributions are summarized as follows:

\begin{itemize}
\item We propose \name, a hybrid transformer and Mamba-based architecture for speech super resolution. We demonstrate that \name outperforms existing state-of-art super resolution methods (U-Net and GAN architectures) by up to 109.1\% on standard intelligibility and quality metrics, with a model size 5.2 million parameters, compared to state-of-art GANs that require at least tens of millions of parameters.

\item We demonstrate that \name is generalizable to multiple acoustic modalities, including over-the-air microphones and bone/vibration-based modalities (BCMs and accelerometers). To adapt \name to contact and vibration-based sensing modalities, we fine-tune \name with only 15 minutes of labeled data collected from the user and demonstrate up to 92.7\% improvements across a variety of different sensor placements, compared to the state-of-art. This is also the first work to sense intelligible speech using only a single head-worn accelerometer.

\item We integrate \name into a wearable and mobile platform to demonstrate real-time speech super resolution and more than a $50$\% reduction in power consumption from sampling and data transmission. Through user studies, we show that vibration-based sensing modalities enhanced by \name generate significantly higher quality speech, with less interference from background noise, than systems that leverage over-the-air microphones with noise reduction algorithms to remove background sounds. 

\item We open-source all our code, designs, and example clips: REDACTED.
\end{itemize}

%% file: sections/related_works.tex
\section{Related Works}
\label{sec:related_works}

\subsection{Audio Super Resolution}

Audio super-resolution, also known as bandwidth expansion, refers to the restoration of high resolution signals from low resolution signals. Specifically, it employs the low frequency components of the low resolution signal to predict and generate the high-frequency components of the signal to improve quality and generate high-resolution audio. Deep learning has shown great promise for audio super-resolution and speech enhancement. U-Net architectures \cite{unet} are one of the most common types of deep learning architectures for audio and speech super resolution, which generally consist of a series of ``contracting'' layers to learn latent features that could be represented with lower dimensionality than speech signals. Then, a series of ``expanding'' layers upsample the embeddings to the final high resolution audio signal, producing high resolution speech. This type of architecture has shown good performance because speech is very structured, with high energy concentrated around specific frequency bands depending on the user's voice and what is spoken.

Kuleshov et al. proposed Audio-UNet \cite{kuleshov2017audio}, the first architecture employing convolutional networks for audio super-resolution tasks. Birnbaum et al. integrated Feature-wise Linear Modulation (FiLM) \cite{perez2018film} with Audio-UNet, resulting in the Temporal FiLM (TFiLM) \cite{birnbaum2019temporal}, which significantly improved the performance of Audio-UNet. Rakotonirina incorporated recurrent neural networks (RNNs) with self-attention \cite{rakotonirina2021self} to further enhance perceptual quality. Other works leverage frequency domain representations as input, such as the three-layer fully connected network proposed by~\cite{li2015deep}. To take advantage of the strengths of both time-domain and frequency-domain processing, several works propose multi-pathway architectures to process both representations simultaneously, such as TFNet~\cite{lim2018time}.

Generative adversarial networks (GANs), such as Seanet \cite{li2021real}, EBEN \cite{hauret2023eben}, and Aero \cite{mandel2023aero}, have recently shown exceptional performance in generating high quality and natural sounding speech from low resolution audio. However, GANs typically employ both a generator and discriminator network, which often exhibits high complexity with numerous associated parameters (on order of tens of billions or more) and long training and fine-tuning times (on order of days to weeks). For these reasons GANs are difficult to incorporate practically into mobile and wearable systems where memory is constrained and real-timeliness is valued.

All of the works mentioned above focus on over-the-air audio super resolution. For speech collected by OTA microphones, it is common for ambient noises to pollute and interfere, often reducing its quality and intelligibility. Moreover, low resolution signals used to train these methods are often generated through low pass filtering clean high resolution speech, before downsampling or decimating. Signals generated in this way are easier for the model to learn because aliasing can be avoided. However, this still requires the system to sample at high frequency. A system looking to leverage the benefits of low resolution signals by reducing the sampling rate to lower power consumption, cannot take advantage of this scheme.


\subsection{Multi-Modal and Vibration-based Speech Enhancement}

Speech enhancement aims to restore high quality speech from degraded speech signals that is often contaminated by noise, echoes, or other interference. The primary objective of speech enhancement is to improve speech clarity, intelligibility, and user satisfaction. Speech super resolution, which aims to reconstruct high frequency formants of a person's voice to improve audio quality, is often viewed as a class of speech enhancement. For OTA audio, ambient background noises can often be heard on the microphone. Another line of speech enhancement focuses on speech denoising or sound source separation to remove background sounds with minimal impact to speech quality~\cite{rethage2018wavenet,lu2013speech,luo2018tasnet,luo2019conv,chatterjee2022clearbuds}. Other works, combine information from multiple sensing modalities, such as a camera or ultrasonic sensors, to extract higher quality speech~\cite{duan2024earse, michelsanti2021overview,hou2018audio,yang2022audio}.

Head-worn bone conduction and vibration-based sensors, such as accelerometers, IMUs, and BCMs, are naturally insensitive to sounds propagating over-the-air. However, high frequency speech components attenuate much more severely through a user's bone and skin than over-the-air. As such, speech enhancement for vibration-based acoustic modalities primarily consists of estimating high frequency formants of speech. Recent works have employed noise-free, but low resolution, speech information from bone conduction sensors or accelerometers in conjunction with high resolution signals from OTA microphones, that may experience interference from ambient sounds, to extract speech and remove background noises. \cite{he2023towards} leverages bone conduction sensors to enhance high resolution signals from OTA microphones, while \cite{wang2022fusing} leverages an IMU to do the same. This work leverages vibration-based sensing modalities, that naturally attenuate background noise, to enhance speech, and focuses on generating and enhancing speech with only a single speech sensing modality at a time.

There are several works that explore generating high quality speech signals using only vibration-based sensors on mobile and wearable platforms~\cite{he2023towards,li2022enabling,maruri2018v}. However, there is a large performance gap between the methods proposed for these platforms (encoder-decoder or U-Net architectures) and state-of-art GANs, which are not as practical to deploy on mobile and wearable platforms. \name addresses this performance gap.


%% file: sections/superresolution.tex
\section{Audio Super Resolution Architecture Design}
\label{sec:superresolution}


\subsection{Opportunities and Challenges}
\label{subsec:background}

\begin{figure}[!t]
    \centering
    \includegraphics[width=0.90\linewidth]{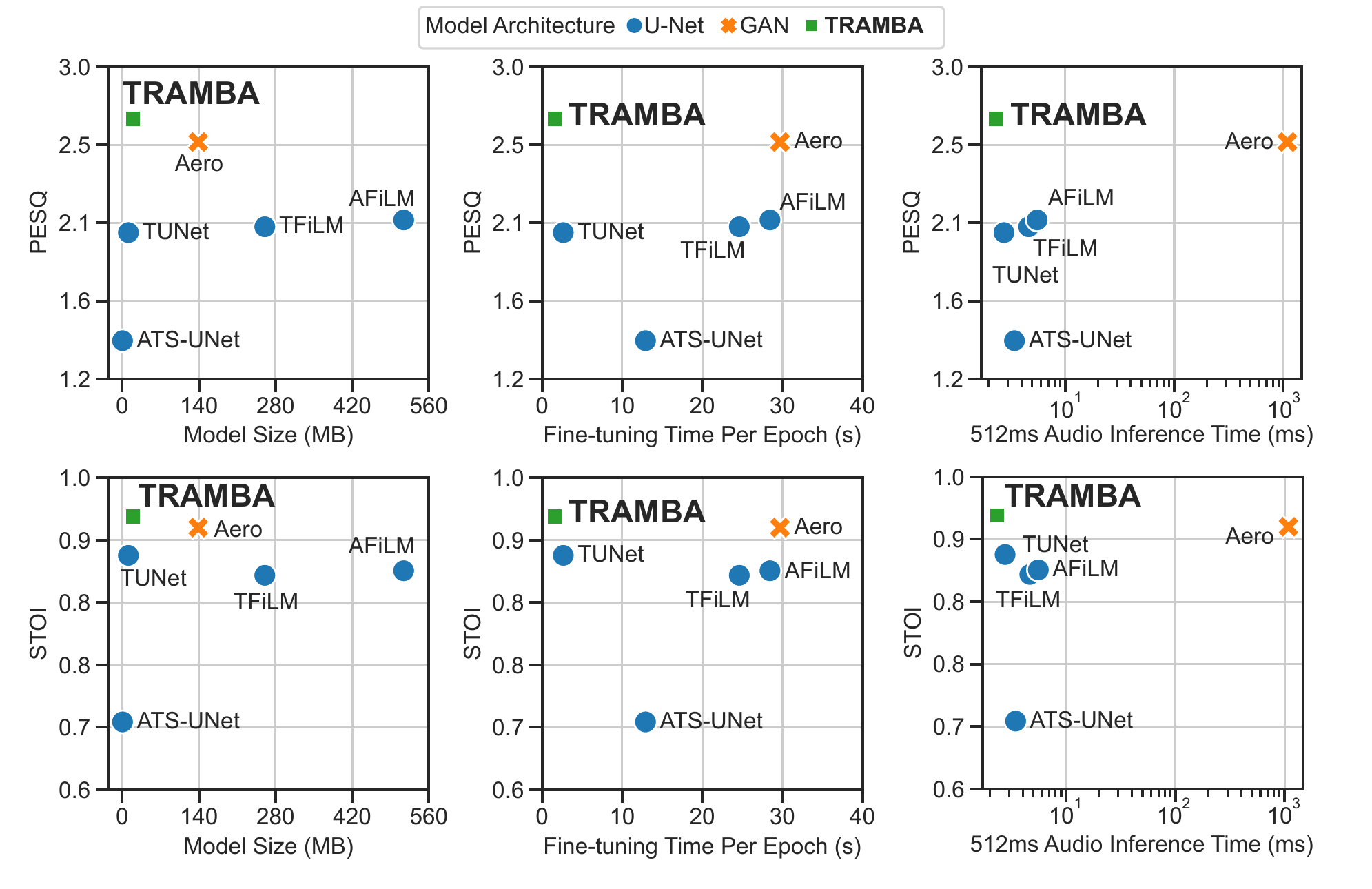}
    \caption{Comparison of performance (PESQ and STOI) vs efficiency (memory footprint, fine-tuning time, inference time) of \name and state-of-art audio super resolution methods.}
    \label{fig:performance_vs_efficiency}
\end{figure}

\begin{figure}[!t]
    \centering
    \includegraphics[width=0.80\linewidth]{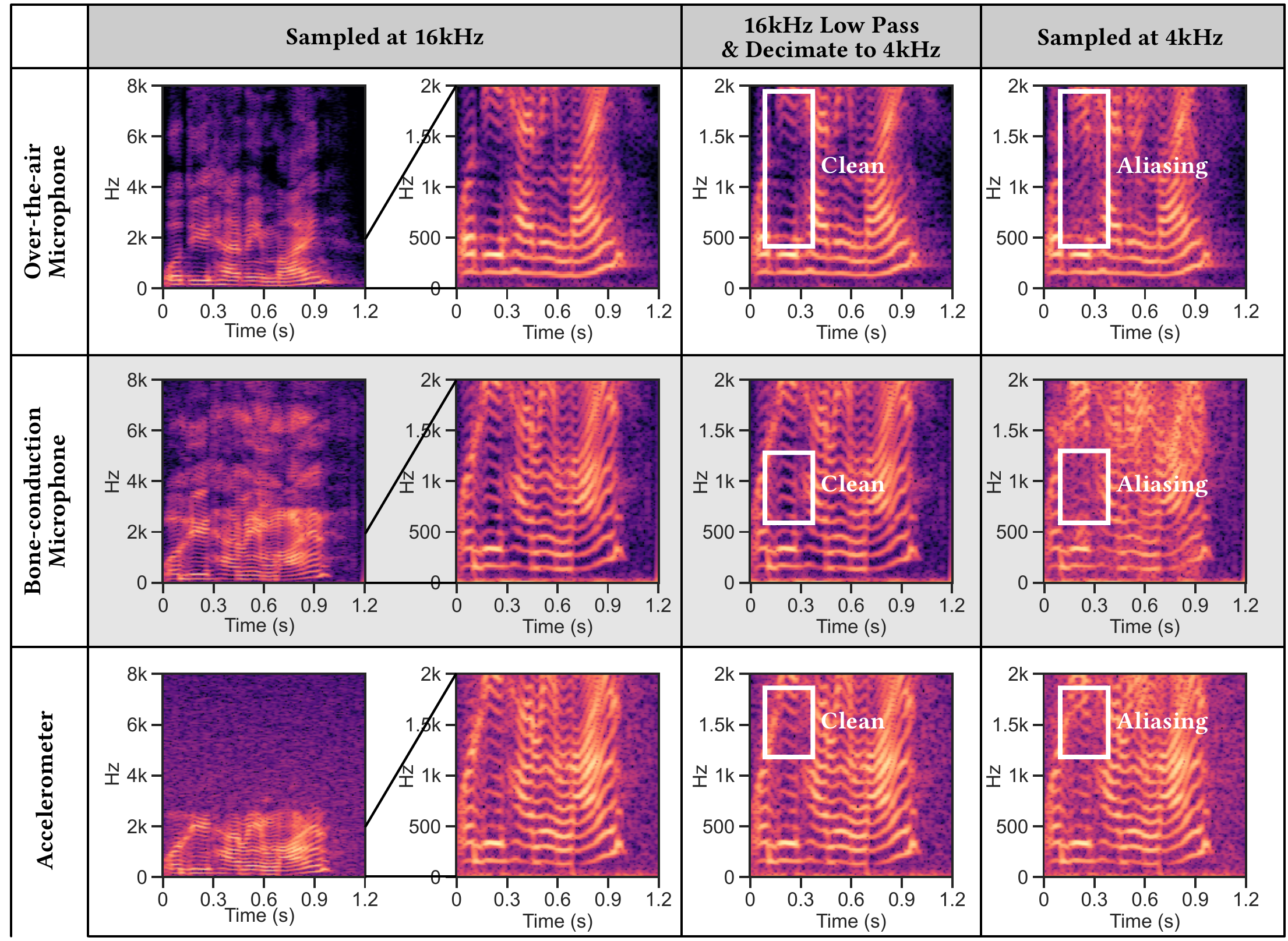}
    \caption{Comparison of OTA microphone, BCM, and accelerometer recorded audio under different sampling and filtering schemes. In the low pass filter + decimate scenario, a $100$ order Butterworth filter with a $2kHz$ cut-off frequency was applied.}
    \label{fig:imu_bcm_audio_visualization}
\end{figure}



Figure~\ref{fig:performance_vs_efficiency} compares the performance and efficiency between \name and six different state-of-art audio super resolution methods, which can broadly be categorized into two approaches: 1) U-Net models and 2) GAN models. U-Net models consist of contracting layers, typically containing convolutional layers, that compress the input into a lower dimensional latent space before expanding back out. This allows the network to learn important low dimensional features common across all training samples to generate higher frequency content in the expansive layers. GANs consist of a generator network that attempts to learn and generate audio waveforms that are indistinguishable from real data by a discriminator network. From Figure~\ref{fig:performance_vs_efficiency}, we see several trends and challenges impeding practical adoption of speech super resolution for bone conduction sensors, which \name addresses.

\subsubsection{Heavy Performance and Performance Gap}
\label{subsubsec:heavy_performance}

The general trend is that the best performing models (typically GANs) require orders of magnitude more memory (tens of billions vs. billions of parameters and hundreds of MBs vs MBs), inference time (hundreds of ms vs. ms), training time (days or weeks vs. minutes), and fine-tuning time than U-Net architectures. However, pure U-Net architectures that are more amenable to resource constrained mobile and wearable platforms see significantly worse performance (PESQ and STOI in Figure~\ref{fig:performance_vs_efficiency}).

\subsubsection{Data Scarcity}
\label{subsubsec:data_scarcity}

Training \name or any bone conduction-based super resolution method requires inputs collected from the BCM or accelerometer, paired with speech collected from OTA microphones that can capture higher frequency formants and act as ground truth. This paired data is not readily available, unlike pure speech or audio collected that can be collected with only a single microphone. Many works that specifically focus on speech enhancement for bone conduction or vibration-based microphones make commendable efforts to collect some amount of data from several volunteers to train/test models~\cite{li2022enabling}, which is hardly enough to train a generalizable model. Other works artificially simulate bone conduction microphones and learn transformations to convert large datasets of audio collected by OTA microphones into signals that you might observe on an IMU or BCM~\cite{he2023towards}, which is used to train the model. However, such methods need to be carefully tuned to generate data that appropriately captures the characteristics of signal observed by an ACCEL or BCM at different sensor locations and different people. Methods trained in this way have difficulty generalizing to new users and sensor placements.

On the other hand, works that focus on over-the-air audio super resolution can train only on standard speech and audio datasets~\cite{hauret2023eben}. These works can easily generate realistic inputs by manually attenuating higher frequency components by applying low pass filters, allowing them to take advantage of the vast amount of publicly available audio. We find that directly applying models trained in this way on OTA speech signals produces poor super resolution results (Section~\ref{subsubsec:unseen_individuals}). 



\subsubsection{System Design Opportunities}
\label{subsubsec:system_design_opportunities}

There are several system design opportunities that we explore, while integrating \name into a mobile and wearable platform.

\noindent
\textbf{Real-Timeliness.} The right column of Figure~\ref{fig:performance_vs_efficiency} shows the performance of each method vs. the inference time on a $512$ms window on an NVIDIA L40 graphics processing unit (GPU). We see that the past best performing models (GANs) require two orders of magnitude longer to perform inference. Some of the models require longer than $512$ms to process the input, meaning it cannot be integrated into a real-time system. \name sees the best performance, while requiring an inference time that is on par with the smallest and fastest running models.

\noindent
\textbf{Power Consumption.} The general trend in energy-efficient systems research is to perform computation as close to the edge as possible (e.g., on the wearable or earable). The outputs in many detection, classification, and recognition tasks is much lower in dimension than the inputs (high resolution raw audio), which significantly reduces the amount of data that needs to be transmitted. Conversely, in this work, we aim to produce high resolution audio (16 kHz). Running the model directly on the wearable would require it to transmit full high resolution audio to the mobile platform, meaning power consumption from wireless transmission cannot be reduced. Moreover, current general purpose microcontrollers (MCU) and system-on-chip (SoC) platforms, that are often used to implement earables and wearables, cannot support models larger than tens or hundreds of KB, which limits the performance of models that can be incorporated ~\cite{max78000,stm32}.


\noindent
\textbf{Sampling Rate.} The left-most column in Figure~\ref{fig:imu_bcm_audio_visualization} shows spectrogram plots of speech recorded from a microphone, BCM, and accelerometer at $16$kHz. We see that most of the higher frequencies above $2$kHz for both the BCM and accelerometer have extremely low energy. This suggests that we can aggressively reduce the sampling rate from the ACCEL and BCM to reduce power consumption of sampling. If model inference and fine-tuning occurs on the mobile platform, the wearable can also reduce the activity of its wireless radio and significantly improve battery life. Moreover, a mobile platform like a smartphone typically has more compute and battery life than a wearable, allowing us to leverage bigger and more powerful models.

However, audio super resolution works operate on audio sampled at the full sampling rate (e.g., 16 kHz), and attempt to fill in higher frequencies~\cite{li2022enabling}. Works that operate on lower sampling frequency inputs tend to low pass filter before decimating the signal. This still requires sampling at a high data rate, which does not improve power consumption on the sensing front. To achieve savings on both sampling and transmission fronts, we aim for robust acoustic super resolution on signals that are directly sampled at a lower frequency rather than filtered signals that have been decimated from higher sampling rates. The right three columns of Figure~\ref{fig:imu_bcm_audio_visualization} shows the difference between these two schemes when decimating to or sampling at $4$kHz. Although the higher frequencies of the BCM and accelerometer. Although the higher frequencies are not as prevalent in vibration-based sensing modalities, aliasing still occurs. The filter + decimate scheme removes aliasing, but is unrealistic in scenarios aiming to reduce power by directly sampling at a lower rate. We show in Section~\ref{subsec:eval_tl_bcm} that \name handles aliasing better than other methods.

\subsection{Deep Neural Network Architecture}
\label{subsec:dnn_architecture}

\begin{figure}[!t]
    \centering
    \includegraphics[width=0.90\linewidth]{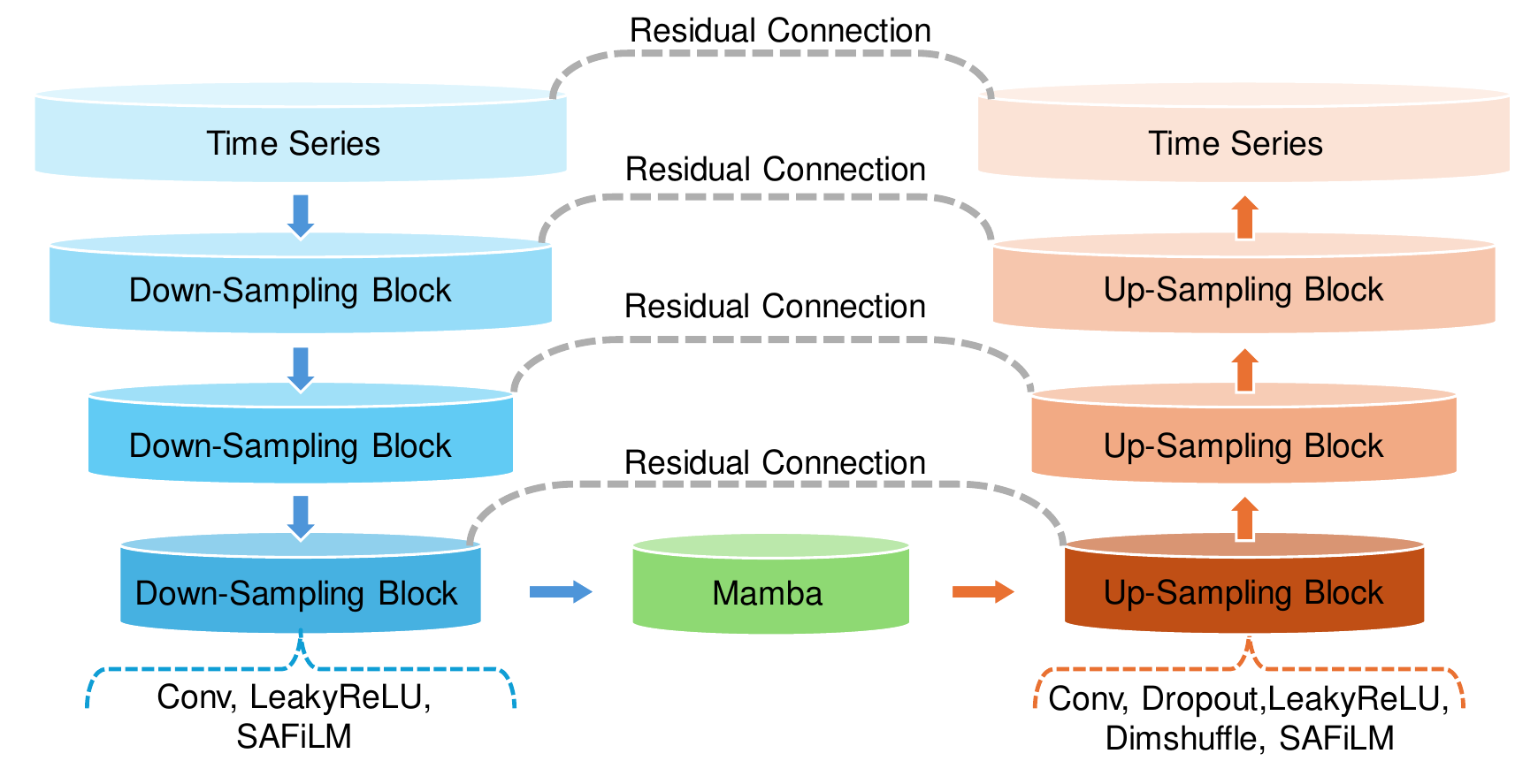}
    \caption{Super resolution and enhancement architecture.}
    \label{fig:dnn_network_architecture}
\end{figure}

Figure~\ref{fig:dnn_network_architecture} shows our super resolution architecture. At a high level our architecture adopts a modified U-Net architecture and incorporates self-attention in the downsampling contracting and expanding layers, as well as Mamba in the narrow bottleneck layer, as shown in Figure~\ref{fig:dnn_network_architecture}.


\subsubsection{Preprocessing}

\name processes $512$ms windows of single-channel audio. An accelerometer measures acceleration across three axes. To preprocess acceleration, we subtract the DC offset such that each axis is zero mean. Then, we average the three axes together to use as input.

\subsubsection{Down-Sampling Block}

Our model comprises a total of three downsampling blocks, each of which contains a 1D convolutional layer that feeds into a layer of LeakyReLU activations, similar to other U-Net models. Each downsampling block contains $2^{5+\textit{b}}$ convolutional filters and a stride of 4. The convolution kernel sizes of each downsampling block are 65, 17, and 7 respectively. Unlike previous works, we apply a novel conditioning layer at the end of each block, called Scale-only Attention-based Feature-wise Linear Modulation (SAFiLM), which we introduce next.


\subsubsection{Scale-only Attention-based Feature-wise Linear Modulation (SAFiLM)}

\begin{figure}[!t]
    \centering
    \includegraphics[width=0.90\linewidth]{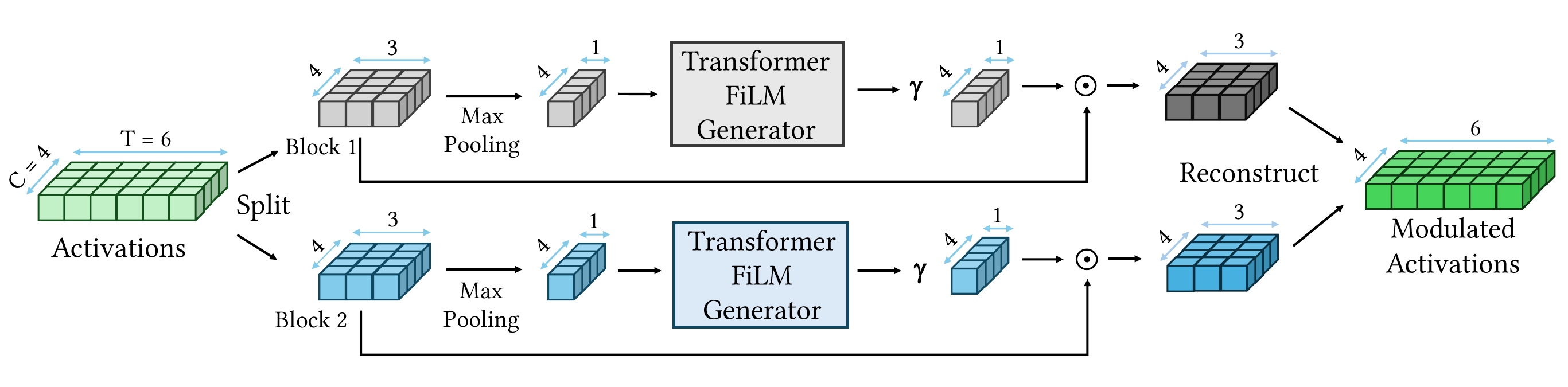}
    \caption{SAFiLM architecture.}
    \label{fig:SAFiLM_architecture}
\end{figure}

General purpose conditioning, which aims to learn an affine transformation consisting of a scaling factor ($\gamma$) and a shift ($\beta$) to apply on intermediate layers in a neural network, has been shown to greatly improve the performance of many tasks. Feature-wise Linear Modulation~\cite{perez2018film} is one of the most commonly used methods, which uses a recurrent neural network (RNN) to learn these parameters. AFiLM~\cite{rakotonirina2021self} replaced the RNN with a transformer block, containing self-attention that better captures long-range dependencies. The transformer's multi-head attention mechanism enables it to learn intricate characteristics of high dimensional input data, while the residual connections and normalization strategies allow gradients to more easily propagate through the layers without vanishing or exploding, which facilitates faster training and better performance even with deeper architectures compared to traditional sequence models like RNNs.

However, transformers are memory intensive. To reduce the memory footprint, SAFiLM (Figure~\ref{fig:SAFiLM_architecture}) only learns and applies the scaling factor, $\gamma$, compared to AFiLM, which learns and applies both a scale and shift across blocks. $T$ and $C$ are the total feature length and number of channels resulting from the 1D convolution. The initial activations \( F \), shaped as \( \mathbb{R}^{T \times C} \), are segmented into \( B \) individual blocks. Each block is represented as \( F_{\text{block}} \) and undergoes dimensionality reduction along its temporal axis via max pooling to form \( F_{\text{pool}} \). Scaling for block b in SAFiLM is expressed as in Equation~\ref{eqn:scaling}. The activations are then modulated as in Equation~\ref{eqn:activation}, where $t$ and $c$ are the feature and channel index resulting from the 1D convolution respectively.

\begin{equation}
    \gamma[b, :] = \text{TransformerBlock}(F_{\text{pool}}[b, :])
    \label{eqn:scaling}
\end{equation}

\begin{equation}
    \text{SAFiLM}(F_{\text{block}}[b, t, c]) = \gamma[b, c] \times F_{\text{block}}[b, t, c]
    \label{eqn:activation}
\end{equation}

\subsubsection{Bottleneck}

Recently, state space sequence models (SSMs) \cite{gu2021combining}, especially structured state space sequence models (S4) \cite{gu2021efficiently}, have shown state-of-art performance in a variety of tasks and modalities, including long sequence time series. The Mamba model \cite{gu2023mamba} builds on top of S4 by incorporating a selection mechanism that allows the model to choose relevant information based on the input, which mimics attention mechanisms found in transformers. This enhancement makes Mamba particularly adept at handling pattern-intensive tasks like language processing. Moreover, Mamba matches the performance of transformers with at least two times less memory. Motivated by these strengths, we decide to incorporate Mamba as our bottleneck rather than a transformer.

Given the benefits of Mamba over transformer, it is natural to replace all transformer blocks (SAFiLM) with Mamba. However, we find that doing so often causes the gradients to vanish, so we keep transformers in the downsampling and upsampling blocks. We plan to explore reasons and solutions for this behavior in future work.

\subsubsection{Up-Sampling Block}

As with the downsampling blocks, our model also comprises three upsampling blocks, each of which performs convolution, dropout, LeakyReLU, pixelshuffle, and SAFiLM. The upsampling block preceding the final output only performs convolution and pixelshuffle. 

We implement a one-dimensional version of \cite{shi2016real}, like in \cite{kuleshov2017audio} for the pixelshuffle layer.

For pixelshuffle layer, we also implement a one-dimensional version of \cite{shi2016real} like \cite{kuleshov2017audio}. If the input of the pixelshuffle layer is ($C_{\text{in}}$, $T_{\text{in}}$) and the scaling factor is 4, the output is ($C_{\text{in}}$ $\div$ 4, $T_{\text{in}}$ $\times$ 4). The three upsampling blocks contain convolution filters of size 512, 256, and 4, with a stride of 4. The convolution kernel sizes of each convolution layer are 7, 17, and 65, respectively.

\subsubsection{Residual Connection}

Many U-Net architectures leverage a skip connection between respective downsample and upsample layers that concatenates the outputs from both. We instead leverage a residual connection, which adds the outputs from both. This allows gradients to more easily flow through layers, facilitating training, reducing the risk of vanishing gradients, and enabling the optimization of deeper networks.


\subsection{Fine-tuning}
\label{subsec:transfer_learning}

As mentioned in Section~\ref{subsec:background}, it is labor intensive to obtain large amounts of paired bone conduction data and clean audio to train a robust model for vibration-based speech enhancement. Rather than take this approach, we instead pre-train our model using the VCTK \cite{CSTRVCTK2019} speech dataset before fine-tuning with a small amount of paired data collected from the user at box opening. This procedure is common in many AI voice generation applications that typically require around $15$ minutes of voice samples from the user~\cite{create2023apple}. To pre-train the model, we downsample clean speech to use as input. During fine-tuning, the clean speech can be captured by the user's mobile phone, while the wearable records from the BCM or ACCEL simultaneously.

%% file: sections/nn_evaluation.tex
\section{Evaluation: Audio Speech Enhancement and Super Resolution}
\label{sec:superresolution_evaluation}

First, we discuss the performance of \name in context of over-the-air speech super resolution. We compare against several state-of-art audio super resolution methods, including four U-Net architectures and two GANs (Section~\ref{subsec:models_compared}).

\subsection{Training, Validation, Testing Procedure and Metrics}
\label{subsec:eval_audio_train_valid_test_procedure_metrics}

\noindent
\textbf{Dataset.} For all methods, we trained, tested, and validated on the VCTK dataset~\cite{CSTRVCTK2019}, which contains a corpus of spoken language comprising 109 native English speakers of various regional accents. Each speaker reads out approximately 400 different sentences, totaling approximately $44$ hours of data. We use audio from 100 individuals for training and the remaining 9 individuals for testing.



\noindent
\textbf{Preprocessing and Training.} For OTA speech enhancement, we downsample high resolution clean audio to $4$kHz to use as input to learn high resolution $16$kHz audio. Training data was divided into smaller segments with a window size of $512$ms and 50\% overlap. For all other methods, we applied the preprocessing steps mentioned the respective papers. We train \name for $30$ epochs. After this time period, we notice that the training, testing, and validation loss changes very little. We trained all other methods using their default training parameters specified in their respective papers and code.

\noindent
\textbf{Loss Function.} In our experiments, we found that using mean squared error (MSE) as the loss function generated lower quality audio than mean absolute error (MAE). As such, we adopted MAE as our primary loss function. However, leveraging MAE, or any type of sample-by-sample distance metric (e.g., MSE), does not necessarily guarantee perceptual quality~\cite{msenot}. As such, we also incorporate multi-resolution STFT loss \cite{yamamoto2020parallel}. We leverage the same multi-resolution STFT loss parameters as \cite{defossez2020real, mandel2023aero}: FFT bins $\in \{512, 1024, 2048\}$, hop lengths $\in \{50, 120, 240\}$, and window sizes $\in \{240, 600, 1200\}$. Our final loss is the sum between MAE and the mutli-resolution STFT loss.


\noindent
\textbf{Performance Metrics.} The primary metrics we use to compare \name for this section and throughout the rest of the paper are discussed next. 

\noindent
\underline{Signal to Noise Ratio (SNR):} SNR is a ratio between the energy of a target signal and background noise, calculated using the formula:

\[
\text{SNR}(x, y) = 10 \log_{10} \left(\frac{\sum x^2}{\sum (x - y)^2}\right)
\]

\noindent
\( x \) represents the magnitude spectrum of the original (high resolution) signal, and \( y \) denotes the magnitude spectrum of the processed signal.

\noindent
\underline{Perceptual Evaluation of Speech Quality (PESQ):} In the context of audio super resolution, PESQ ~\cite{pesq} serves as an objective measurement to evaluate the quality of reconstructed high-resolution speech signals as perceived by human listeners. This metric compares the generated audio against a high-quality reference, with scores typically ranging from -0.5 to 4.5. A higher PESQ score indicates a closer distance to the original audio's quality.

\noindent
\underline{Log-Spectral Distance (LSD):} This metric measures the distance between the spectrum of the generated speech and the clean speech on the log scale. The difference is measured between log spectra because human hearing follows this scale~\cite{lsd}. Scores are calculated over all frequency bins, with lower values indicating a closer match to the reference spectrum, as shown below.


\[
\text{LSD}(x, y) = \frac{1}{T} \sum_{t=1}^T \sqrt{\frac{1}{K} \sum_{k=1}^K \left(\log X(t, k) - \log \hat{X}(t, k)\right)^2}
\]

\noindent
$X(t, k)$ denotes the magnitude spectrum of the original signal at time frame $t$ and frequency $k$. $\hat{X}(t, k)$ represents the magnitude spectrum of the super-resolved signal at the same time frame and frequency. $T$ indicates the total number of time frames, and $K$ is the total number of frequency bins.

\noindent
\underline{Short-Time Objective Intelligibility (STOI):} STOI \cite{stoi} compares the clarity and intelligibility of speech enhanced from a lower to a higher resolution against a clean reference, with scores ranging from 0 to 1. A higher STOI score, closer to 1, signifies better intelligibility, indicating that the enhanced speech is easier to understand.


\subsection{Models Compared}
\label{subsec:models_compared}

We compare against six additional models. Four of the models are UNet \cite{unet} structures, while the remaining two employ GAN \cite{gan} structures. 

\begin{itemize}
\item TFiLM~\cite{birnbaum2019temporal} is a UNet model that uses a recurrent neural network to alter the activations of the convolutional layers in the standard U-Net. 

\item AFiLM \cite{rakotonirina2021self} replaces the RNNs in TFiLM with self-attention. 

\item TUNet \cite{nguyen2022tunet} uses the Performer \cite{performer} achitecture as the bottleneck of the UNet model.

\item ATS-UNet \cite{li2022enabling} is a lightweight UNet model that reduces the number of parameters in the convolutional layers. 

\item EBEN \cite{hauret2023eben} is a GAN for audio speech super resolution with a lightweight generator model. 

\item Aero \cite{mandel2023aero} is a GAN that operates directly on the complex-valued frequency representation of audio. It employs two independent channels to process amplitude and phase information, circumventing the Spectral discontinuity and phase mismatch associated with high and low frequency splicing that are inherent to traditional methods.
\end{itemize}

\subsection{Performance Summary}
\label{subsec:eval_audio_performance_summary}

\begin{table}[ht]
  \centering
  
  \begin{tabular}{|c|c|c|c|c|c|c|c|} 
    \hline
     \multirow{ 2}{*}{}& \multirow{ 2}{*}{Params (M)} & Model & Inference & \multirow{ 2}{*}{SNR} & \multirow{ 2}{*}{PESQ} & \multirow{ 2}{*}{STOI} & \multirow{ 2}{*}{LSD} \\ 
     & & Size (MB) & Time (ms) & & & & \\ \hline
     TFiLM~\cite{birnbaum2019temporal}& 68.2& \ 260.3& 4.6423&  13.7079&2.4349& 0.8475& 1.8667\\ 
     AFiLM~\cite{rakotonirina2021self}& \ 134.7& \ 513.9& 5.5552&  11.3508&2.4658& 0.8689& 1.8665\\ 
     TUNet~\cite{nguyen2022tunet}& \ 2.9& \ 11.2& 2.7550&  22.3046&2.5783& 0.9382& 0.9429\\ 
     ATS-UNet~\cite{li2022enabling}& 0.1& \ 0.5& 3.4419&  12.1807& 1.5471& 0.6695& 1.5624 \\ 
     EBEN~\cite{hauret2023eben}& \ 29.7& \ 113.3& 242.1341&  18.9196&2.5800& 0.8912& 1.0079\\ 
     Aero~\cite{mandel2023aero}& \ 36.3& \ 138.7& 1084.5392&  22.5075&3.0136& 0.9386& 0.8102\\
     \textbf{\name}& \ \textbf{5.2}& \ \textbf{19.7}& \textbf{2.3309}&  \textbf{23.2426}&\textbf{3.2344}& \textbf{0.9478}& \textbf{0.8271}\\ \hline
  \end{tabular}

  \caption{Summary of performance of \name compared to existing audio super resolution methods. Inference time is measured on an NVIDIA L40 GPU processing one window of 512 ms.}
  \label{tab:superresolution_performance_summary}
\end{table}

Table~\ref{tab:superresolution_performance_summary} summarizes the test performance across all methods. We see that \name (highlighted) has the best performance across all metrics and sampling rates, while requiring similar processing times and memory as U-Net models. The Aero GAN method slightly outperforms \name in the LSD metric. LSD measures signal similarity more than perceptual quality and \name outperforms Aero in all other perceptual and noise metrics (SNR, PESQ, STOI). This demonstrates the addition of transformers and Mamba improved the modeling and generation of local speech formants than traditional U-Net architectures and encoder-decoder architectures. This also demonstrates that transformer and Mamba-based architectures can outperform state-of-art GANs with only a fraction of the memory and inference time. Both GANs require significantly more time to perform inference on an L40 GPU. Aero requires more time than the length of the window size it processes ($1080$ms vs. $512$ms), meaning it cannot run in real-time. For these reasons, we compare with the best performing U-Net architecture (TUNet) for the rest of this section.


\subsection{Reducing Sampling Rate}
\label{subsec:reducing_sampling_rate}

\begin{figure}[!t]
    \centering
    \resizebox{0.8\linewidth}{!}{%
        \begin{subfigure}[b]{0.30\linewidth}
            \centering
            \includegraphics[width=\linewidth]{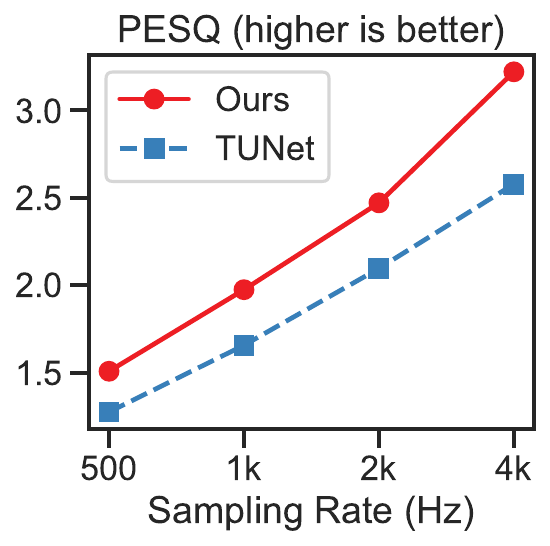}
        \end{subfigure}
        \hfill 
        \begin{subfigure}[b]{0.30\linewidth}
            \centering
            \includegraphics[width=\linewidth]{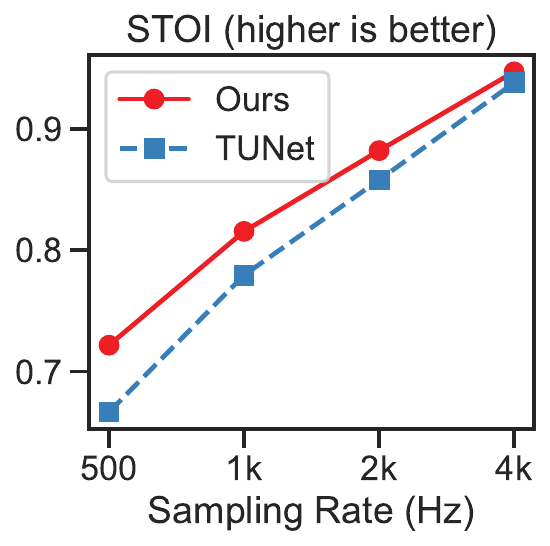}
        \end{subfigure}
        \hfill 
        \begin{subfigure}[b]{0.30\linewidth}
            \centering
            \includegraphics[width=\linewidth]{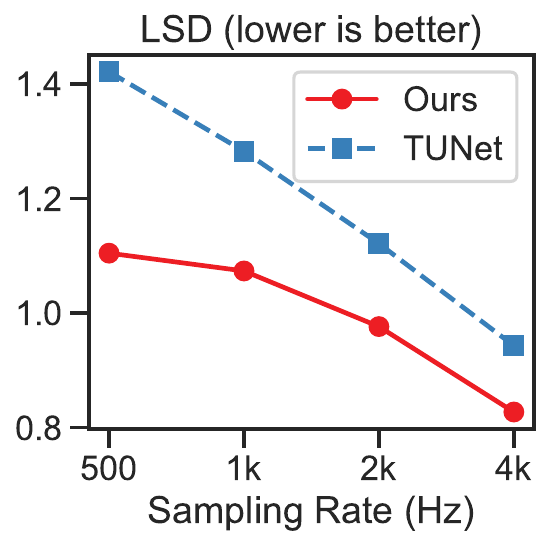}
        \end{subfigure}
    } 
    \caption{Effects of the sampling rate on audio speech super resolution.}
    \label{fig:samplerate_superresolution}
\end{figure}

Figure~\ref{fig:samplerate_superresolution} shows the super resolution performance of \name to $16$kHz as the sampling rate of the input signal varies from 500Hz to 4kHz. Even as the sampling rate decreases down to 500Hz, \name outperforms TUNet. Even if the sampling rate decreases down to $2$kHz or $1$kHz, \name achieves comparable performance to other models at higher sampling rates.

\subsection{Super Resolution Architecture Ablation Study}
\label{subsec:architecture_ablation Study}

\begin{table}[ht]
  \centering
  \begin{tabular}{|c|c|c|c|c|c|} 
    \hline
    & PESQ& STOI& LSD& Parameter & Train Time Per Epoch \\ \hline
    Replace Mamba with Performer & 3.1208 & 0.9474 & 0.8042 & 6.5M & 243.93s \\ 
    Remove SAFiLM & 2.8056 & 0.9428 & 0.8268 & 2.6M & 114.04s \\ \hline
    TRAMBA& 3.2344& 0.9478 & 0.8271 & 5.2M & 191.61s \\ \hline
  \end{tabular}

  \caption{Ablation study.}
  \label{tab:architecture_ablation_study}
\end{table}

Figure~\ref{tab:architecture_ablation_study} shows the results of our ablation study, which explores the impact of combining self-attention, transformers, and Mamba on speech enhancement. Replacing Mamba in our bottleneck layer with Performer, a variant of transformers, yielded similar performance metrics (PESQ, STOI, LSD). However, the training time increased significantly from 191.61 seconds to 243.93 seconds per epoch and the number of parameters increased to $6.5$M. Mamba not only maintains performance but also significantly increases model training speed and reduces the required number of model parameters.

Next, we removed attention (SAFiLM) from the contracting and expansion layers and observed that the training time per epoch decreased by more than $50$\%. However, the LSD, STOI, and particulary PESQ dropped significantly. This performance loss highlights the importance of transformers and self-attention in enabling models to learn and generate complex patterns of speech. The results of our ablation study demonstrates the importance of both self-attention and Mamba in creating a robust model for generating high quality speech, while being efficient in time and memory footprint.

\section{Evaluation: Finetuning from Audio Super Resolution to ACCEL and BCM Speech Enhancement}
\label{sec:eval_transfer_learning}

In this section, we discuss the performance of \name on enhancing speech captured by accelerometer readings and bone conduction microphones on areas of the face where (future) wearables are commonly worn. As discussed in Section~\ref{subsec:background} collecting ground truth pairs (ACCEL/BCM, audio) is extremely labor intensive. As such, we onlt fine-tune our over-the-air audio speech super resolution model (Section~\ref{sec:superresolution_evaluation}) with a small amount of paired data collected from an individual, which is similar in procedure to many existing AI voice generation applications~\cite{create2023apple}.

\subsection{Data Collection}
\label{subsec:eval_tl_data_collection}

\begin{figure}[!t]
    \centering
    \includegraphics[width=0.90\linewidth]{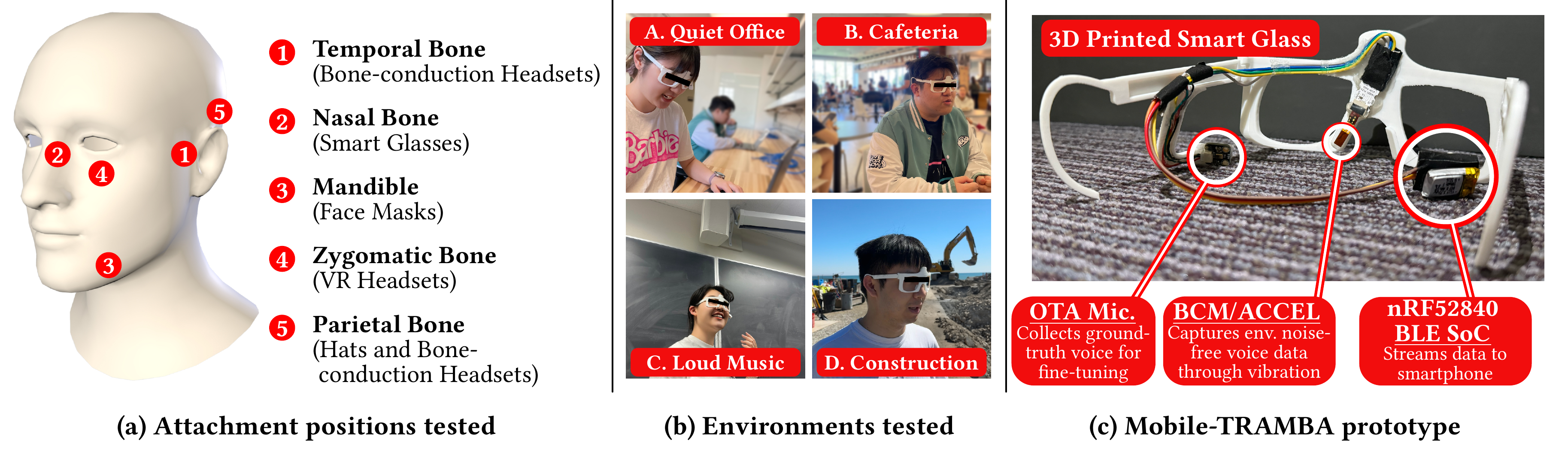}
    \caption{Data collection setup, environments deployed, and wearable prototype.}
    \label{fig:datacollection_setup}
\end{figure}

We recruit 7 volunteers (4 males, 3 females) between 18-30 years old. All studies were approved by the \textit{REDACTED} IRB. We collect data from each volunteer, placing the ACCEL or BCM at each location marked in Figure~\ref{fig:datacollection_setup}a. At each location, we collect a total of approximately 20 minutes of data. We instruct each volunteer to read a set of sentences from an e-book on basic spoken English~\cite{BasicSpokenEnglish}. We vary the amount of data used to fine-tune \name to analyze the effects of the quantity of data on performance.

Figure~\ref{fig:datacollection_setup}c shows the glasses wearable we created to both collect data and deploy in various environments. There is a holder for the ACCEL/BCM to collect inputs and a holder for an over-the-air microphone to capture high resolution ground truth speech. For the BCM, we use the V2S200D voice vibration sensor~\cite{v2s2024knowles} and the IIM-42352 for the MEMS accelerometer~\cite{iim2021invensense}.

\subsection{Effects of Preprocessing: Downsampling}
\label{subsec:effects_preprocessing}

\begin{figure}[!t]
    \centering
    \includegraphics[width=0.90\linewidth]{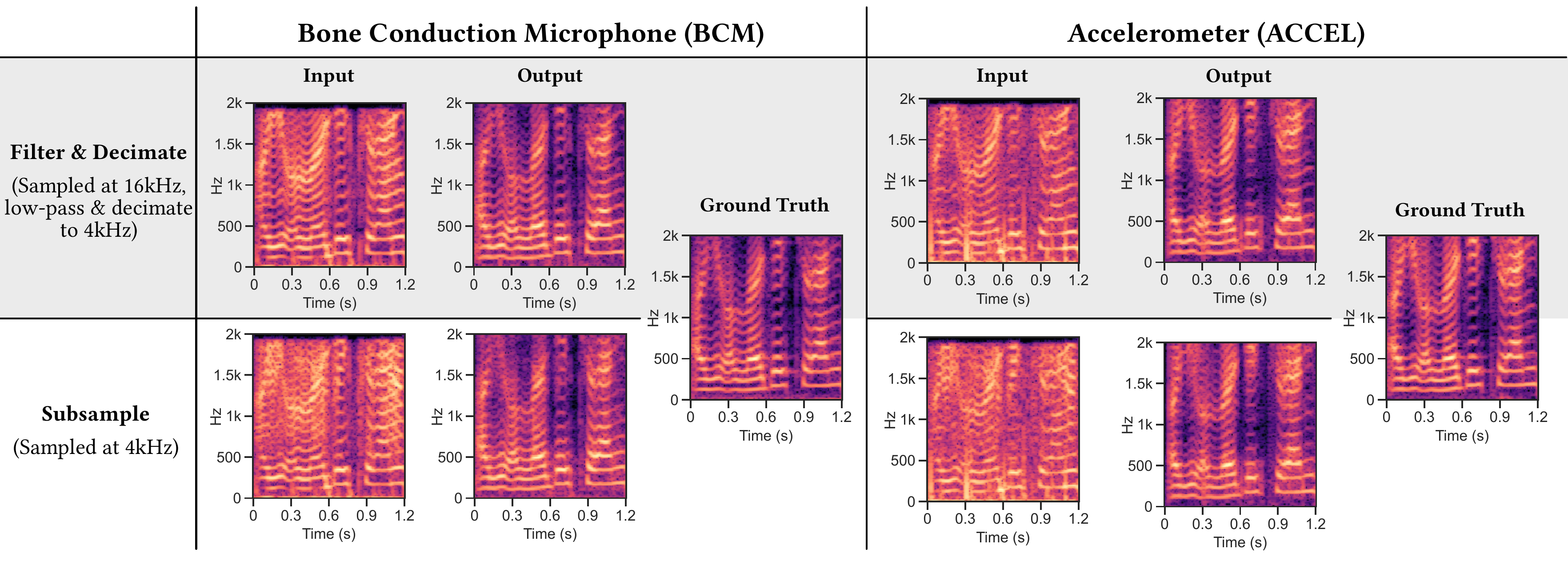}
    \caption{Effects of downsampling.}
    \label{fig:effects_downsampling}
\end{figure}

\begin{figure}[!t]
    \centering
    \includegraphics[width=0.90\linewidth]{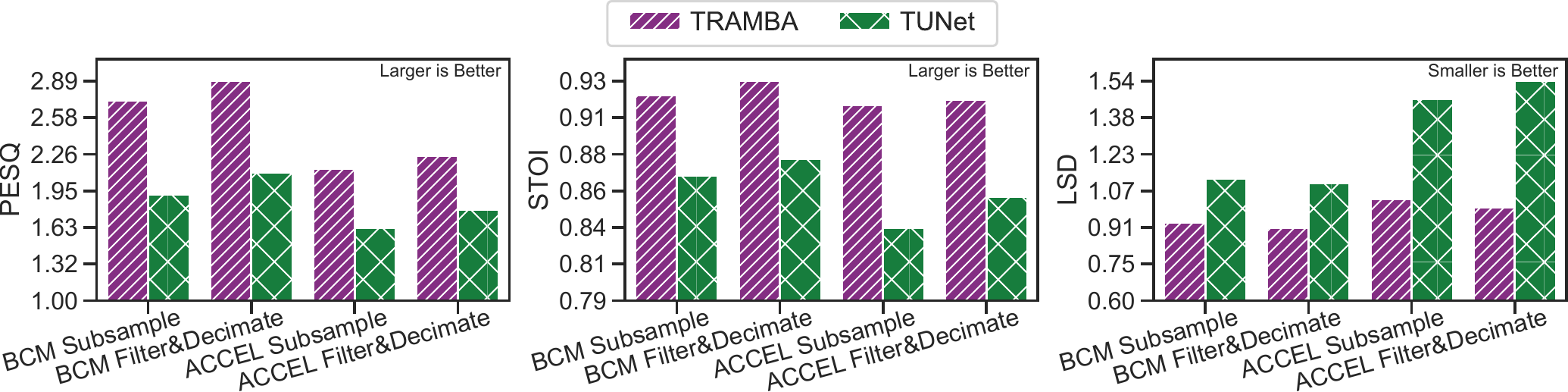}
    \caption{Downsampling effects on performance.}
    \label{fig:effects_downsampling_numbers}
\end{figure}

As discussed in Section~\ref{subsec:background}, most works that develop super resolution methods for both over-the-air and vibration-based speech sensing 1) still sample at the non-decimated sampling rate (e.g., $16$kHz), or 2) perform low pass filtering before subsampling. In both scenarios, aliasing is avoided or significantly reduced, but we cannot reduce power consumption by sampling at a lower rate. As such, we directly 3) decimate speech signals, without any filtering, for an accurate representation of a slower sampling rate.

Figure~\ref{fig:effects_downsampling} and Figure~\ref{fig:effects_downsampling_numbers} compares the performance of \name after downsampling from $16$kHz to $4$kHz with the two downsampling approaches, 2) and 3). For method 2), we apply a low pass filter with a cut-off frequency of $2$kHz before subsampling. We see that \name outperforms other methods across all metrics. Although there is a drop in performance for \name if we remove filtering before decimation (3), \name still outperforms most other methods even if the other methods filter before decimating (2). Additionally, the performance drop-off is significantly steeper for other methods. For the rest of this section, we directly decimate audio signals, 3).

\subsection{BCM and ACCEL-based Speech Enhancement}
\label{subsec:eval_tl_bcm}

\begin{table*}[!t]
	\centering
	\resizebox{1\textwidth}{!}{
		\begin{tabular}{|c|c|c|c|c|c|c|c|c|c|}
              \cline{3-10}
              \multicolumn{2}{c}{} & \multicolumn{4}{|c||}{\textbf{BCM}}& \multicolumn{4}{c|}{\textbf{ACCEL}} \\
		   \cline{3-10}
              \hline
			 Method & Fine-tuning Time Per Epoch & PESQ & STOI & LSD & \multicolumn{1}{c||}{WER} & PESQ & STOI & LSD & WER \\
             \hline

             TFiLM & 24.62s & 2.1390 & 0.8769 & 1.3342 & \multicolumn{1}{c||}{8.73\%} & 1.6577 & 0.8486 & 1.4760 & 12.89\% \\
             AFiLM & 28.45s & 2.2618 & 0.8784 & 1.4657 & \multicolumn{1}{c||}{6.48\%} & 1.6220 & 0.8422 & 1.7090 & 15.66\% \\
             TUNet & 2.65s & 1.9195 & 0.8701 & 1.1220 & \multicolumn{1}{c||}{6.21\%} & 1.6337 & 0.8366 & 1.4649 & 13.29\% \\
             ATS-UNet & 12.91s & 1.4155 & 0.7398 & 1.5370 & \multicolumn{1}{c||}{29.86\%} & 1.2981 & 0.6356 & 1.5979 & 71.80\% \\
             Aero & 29.70s & 2.6669 & 0.9072 & 0.8917 & \multicolumn{1}{c||}{4.48\%} & 2.0172 & 0.8993 & 1.1009 & 14.04\% \\
             \textbf{\name} & \textbf{1.58s} & \textbf{2.7273} & \textbf{0.9221} & \textbf{0.9329} & \multicolumn{1}{c||}{\textbf{2.65\%}} & \textbf{2.1365} & \textbf{0.9159} & \textbf{1.0326} & \textbf{7.76\%} \\
             \hline
		\end{tabular}
	}

 \caption{Performance summary of speech enhancement from $4$kHz to $16$kHz. Models were fine-tuned using $15$ minutes of data collected at Location 2 (Nasal Bone) for 52 minutes. Models were fine-tuned on an NVIDIA L40 GPU.}
 \label{tbl:eval_bcm_imu_summary}
\end{table*}

\begin{figure}[!t]
    \centering
        \begin{subfigure}[b]{0.30\linewidth}
            \centering
            \includegraphics[width=\linewidth]{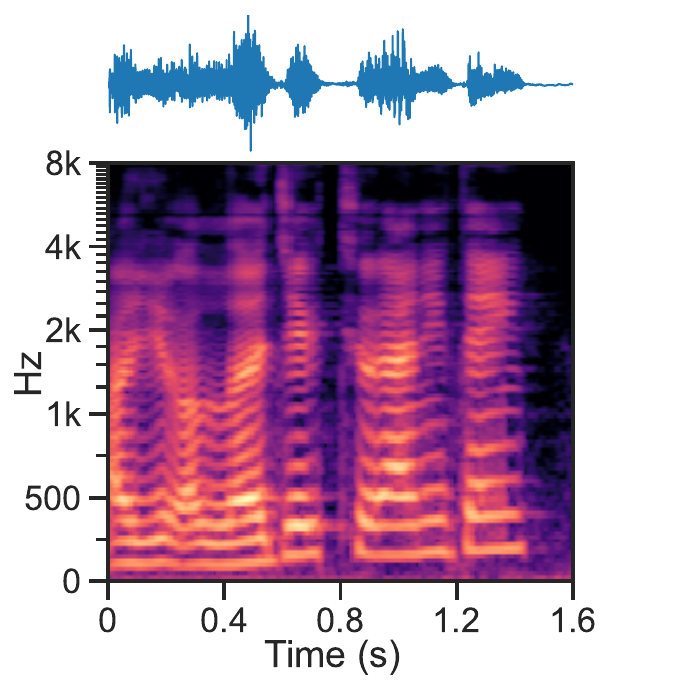}
            \caption{Ground Truth}
        \end{subfigure}
        \hfill
        \begin{subfigure}[b]{0.30\linewidth}
            \centering
            \includegraphics[width=\linewidth]{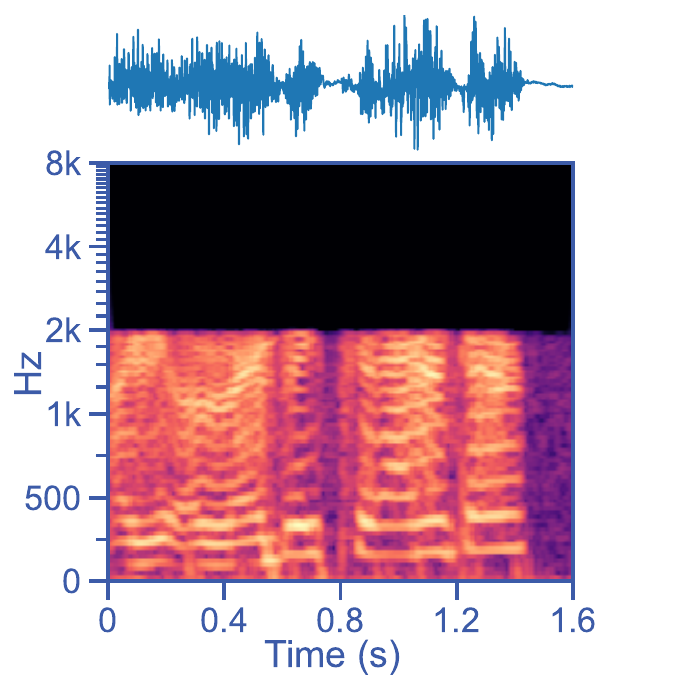}
            \caption{BCM Input (4kHz)}
        \end{subfigure}
        \hfill
        \begin{subfigure}[b]{0.30\linewidth}
            \centering
            \includegraphics[width=\linewidth]{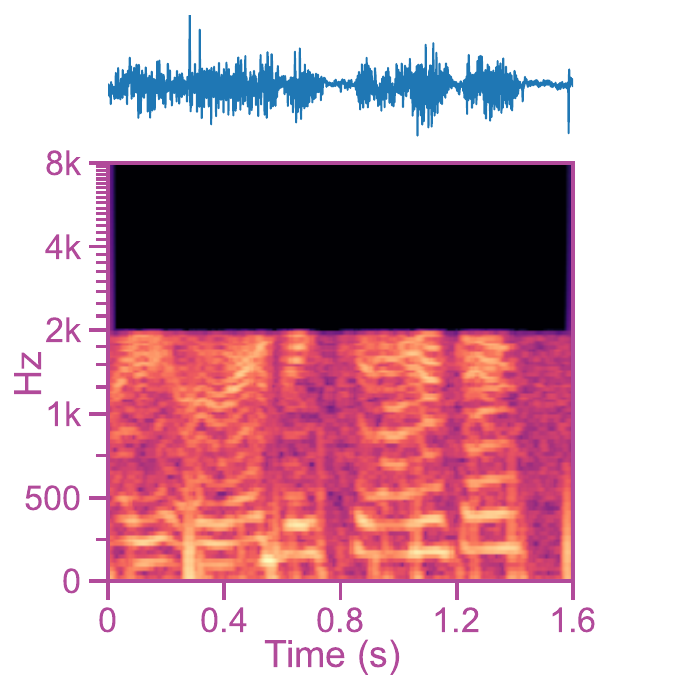}
            \caption{ACCEL Input (4kHz)}
        \end{subfigure}
        \vskip\baselineskip
        \begin{subfigure}[b]{0.30\linewidth}
            \centering
            \includegraphics[width=\linewidth]{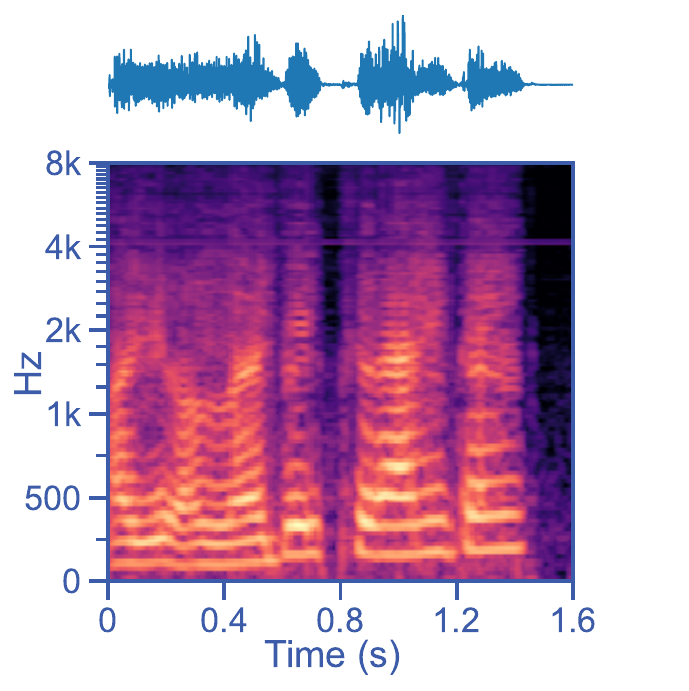}
            \caption{BCM - TUNet (UNet)}
        \end{subfigure}
        \hfill
        \begin{subfigure}[b]{0.30\linewidth}
            \centering
            \includegraphics[width=\linewidth]{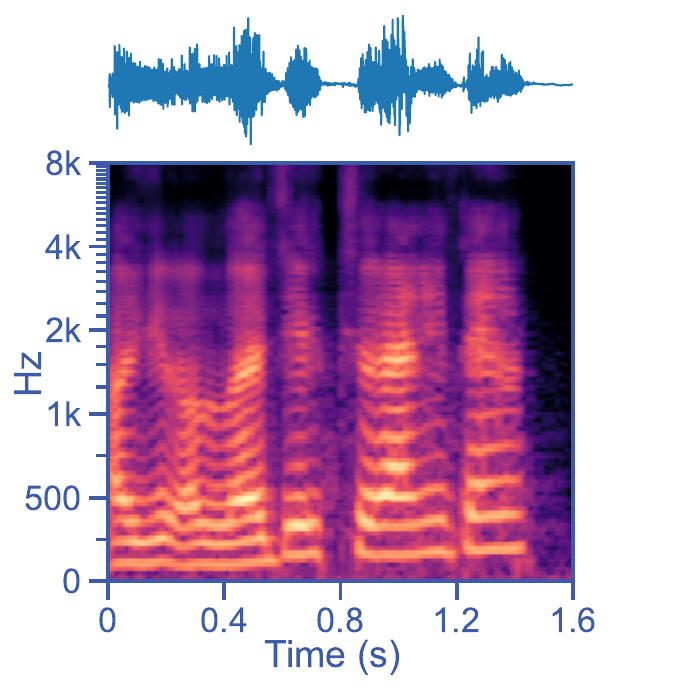}
            \caption{BCM - Aero (GAN)}
        \end{subfigure}
        \hfill
        \begin{subfigure}[b]{0.30\linewidth}
            \centering
            \includegraphics[width=\linewidth]{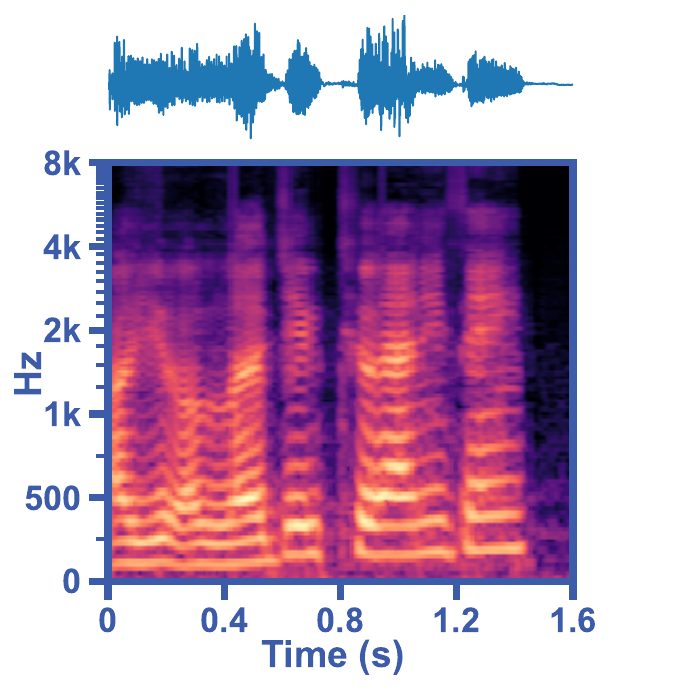}
            \caption{\textbf{BCM - \name}}
        \end{subfigure}
        \vskip\baselineskip
        \begin{subfigure}[b]{0.30\linewidth}
            \centering
            \includegraphics[width=\linewidth]{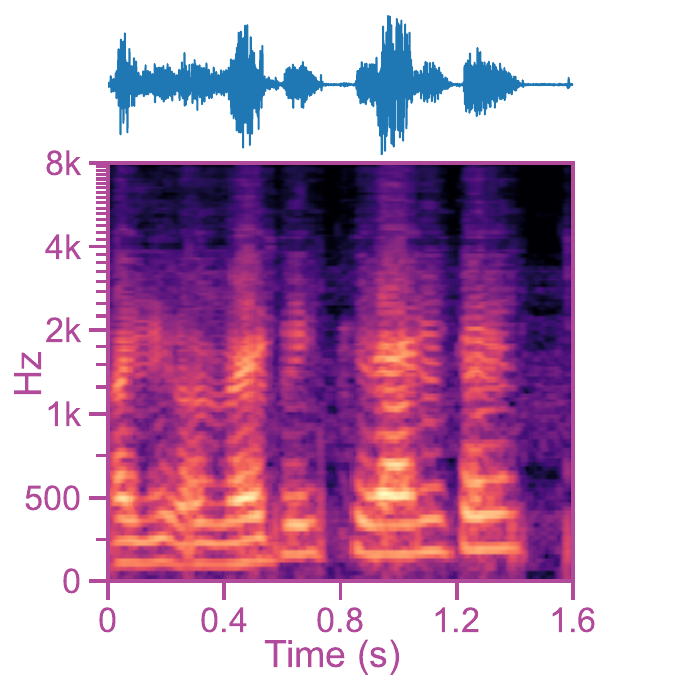}
            \caption{ACCEL - TUNet (UNet)}
        \end{subfigure}
        \hfill
        \begin{subfigure}[b]{0.30\linewidth}
            \centering
            \includegraphics[width=\linewidth]{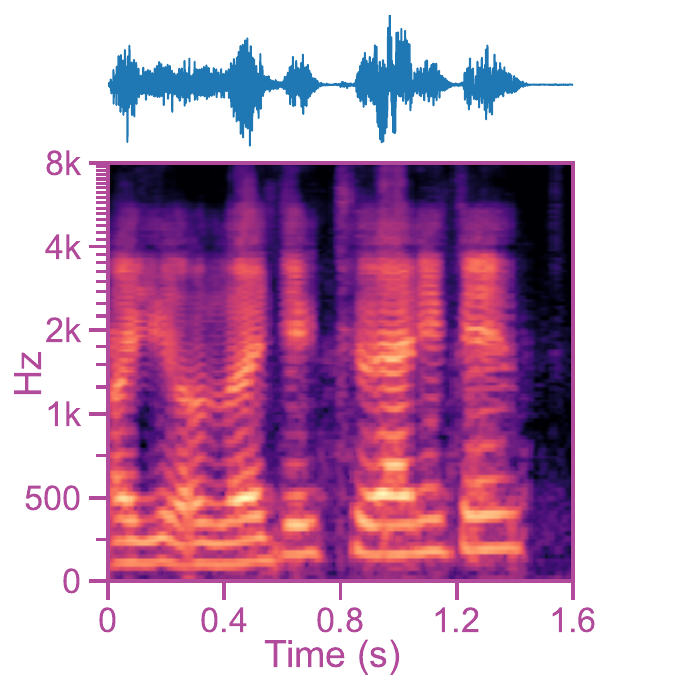}
            \caption{ACCEL - Aero (GAN)}
        \end{subfigure}
        \hfill
        \begin{subfigure}[b]{0.30\linewidth}
            \centering
            \includegraphics[width=\linewidth]{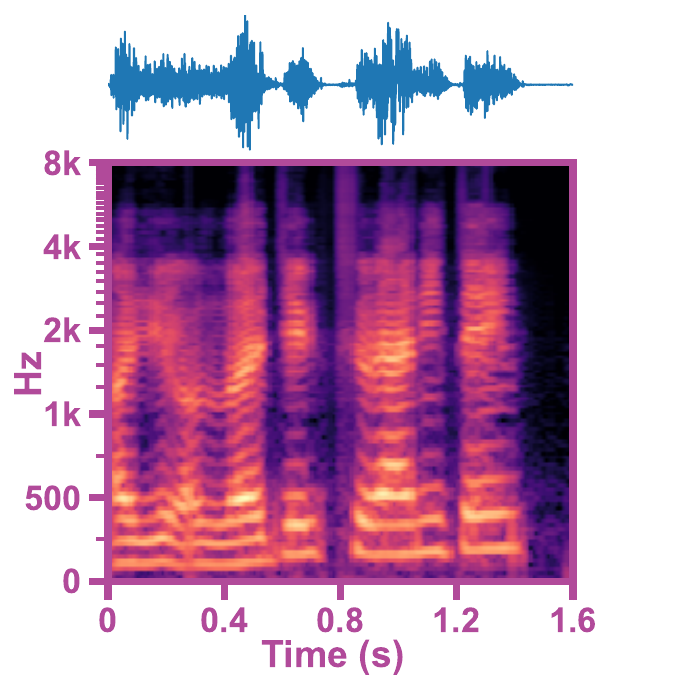}
            \caption{\textbf{ACCEL - \name}}
        \end{subfigure}
    \caption{Visualizing reconstructed audio from ACCEL and BCM after finetuning BCM/ACCEL + audio data pairs. The BCM/ACCEL is placed at location 2 (Nasal Bone). (b) and (c) show no energy at frequencies higher than $2$kHz because the BCM and ACCEL were sampled at $4$kHz.}
    \label{fig:transfer_visualization_bcm_imu_audio}
\end{figure}

Table~\ref{tbl:eval_bcm_imu_summary} summarizes the performance of \name for speech super resolution on BCMs and ACCELs, subsampled at $4$kHz to $16$kHz sampling rate that most works benchmark. The values here represent the average performance across all participants and BCM/ACCEL placements. The models were fine-tuned on each volunteer with $15$ minutes of data at Location 2 (Nasal Bone) as shown in Figure \ref{fig:datacollection_setup} for 52 minutes. We see that \name significantly outperforms existing methods for both sensing modalities and across all metrics. 

One additional metric we compare is the word error rate (WER), which is the ratio between the number of incorrect words perceived and the total number of spoken words. To compute this metric, we input all generated speech signals into Whisper~\cite{whisper}, a state-of-the-art automatic speech recognition (ASR) model and compare with ground truth transcript.

Compared to the larger GAN and U-Net methods, \name requires only a fraction of the time to fine-tune (1.6 seconds vs 30 seconds per epoch). Figure~\ref{fig:transfer_visualization_bcm_imu_audio} visualizes the BCM and accelerometer reconstruction result of one speech segment compared to the best performing GAN (Aero) and U-Net (TUNet). Both \name and Aero clearly generate higher quality results than TUNet. \name rectifies aliased speech better than the GAN, as evidenced at the end of the segment. The formants in the original clean speech stays level or trends upwards. However, the BCM and ACCEL are both aliased when sampled at $4$kHz. Aero reconstructs the end of the segment with the formants trending downward in the higher frequencies due to aliasing. However, \name reconstructs speech much more closely to the original speech pattern. 

Next, we explore how \name performs depending on sensor placement, across different volunteers, and availability of labeled data. For the rest of the section, we compare \name against method TUNet, as it was the second best performing method overall outside of the GANs. While the GAN architectures generally outperformed the U-Net architectures, they take orders of magnitude longer to train, fine-tune, and perform inference, making them impractical to deploy.

\subsubsection{BCM and ACCEL Placement}
\label{subsubsec:placement}

\begin{figure}[!t]
    \centering
    \includegraphics[width=\linewidth]{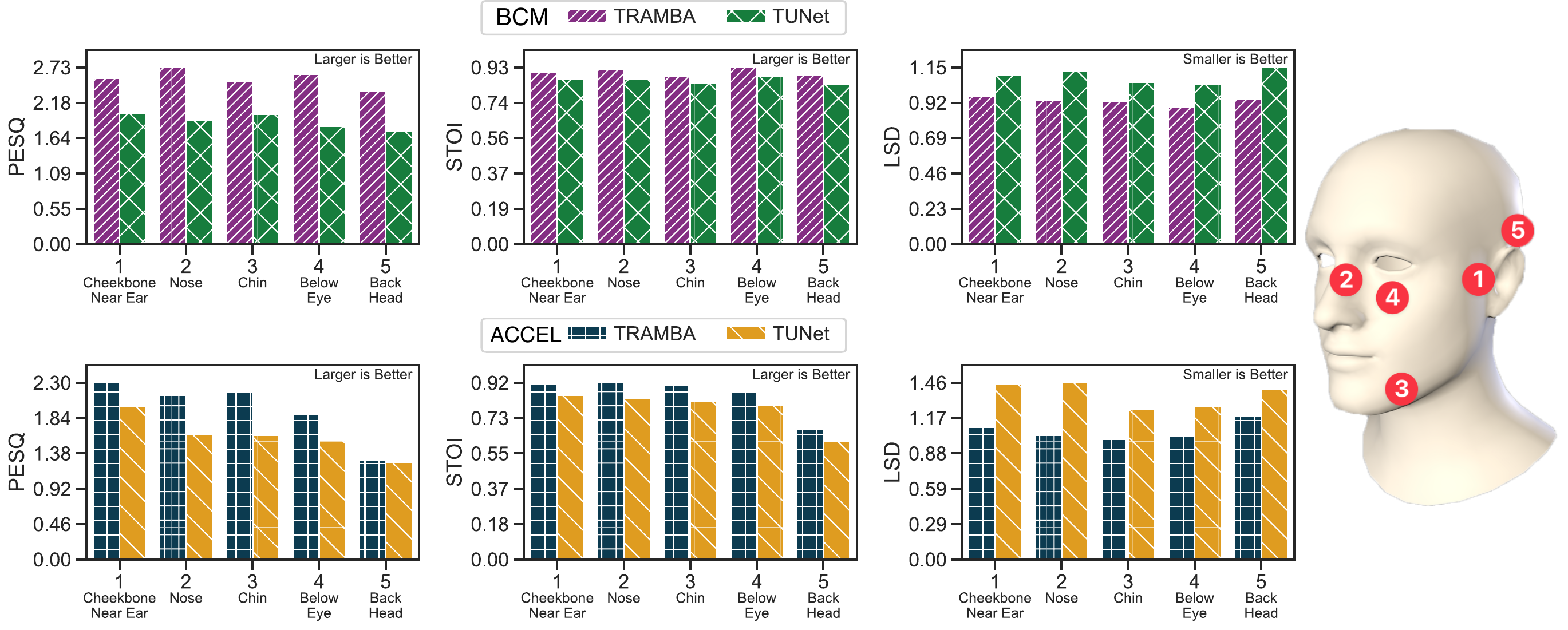}
    \caption{Performance of BCM/ACCEL speech enhancement at different locations.}
    \label{fig:transfer_bcm_finetuning_locations}
\end{figure}

Figure~\ref{fig:transfer_bcm_finetuning_locations} illustrates the performance breakdown of \name by BCM and ACCEL placement location. \name has consistent and better performance across all common and practical locations for a BCM or ACCEL.

\subsubsection{Breakdown Across Individuals}
\label{subsubsec:individuals}

\begin{figure}[!t]
    \centering
    \includegraphics[width=\linewidth]{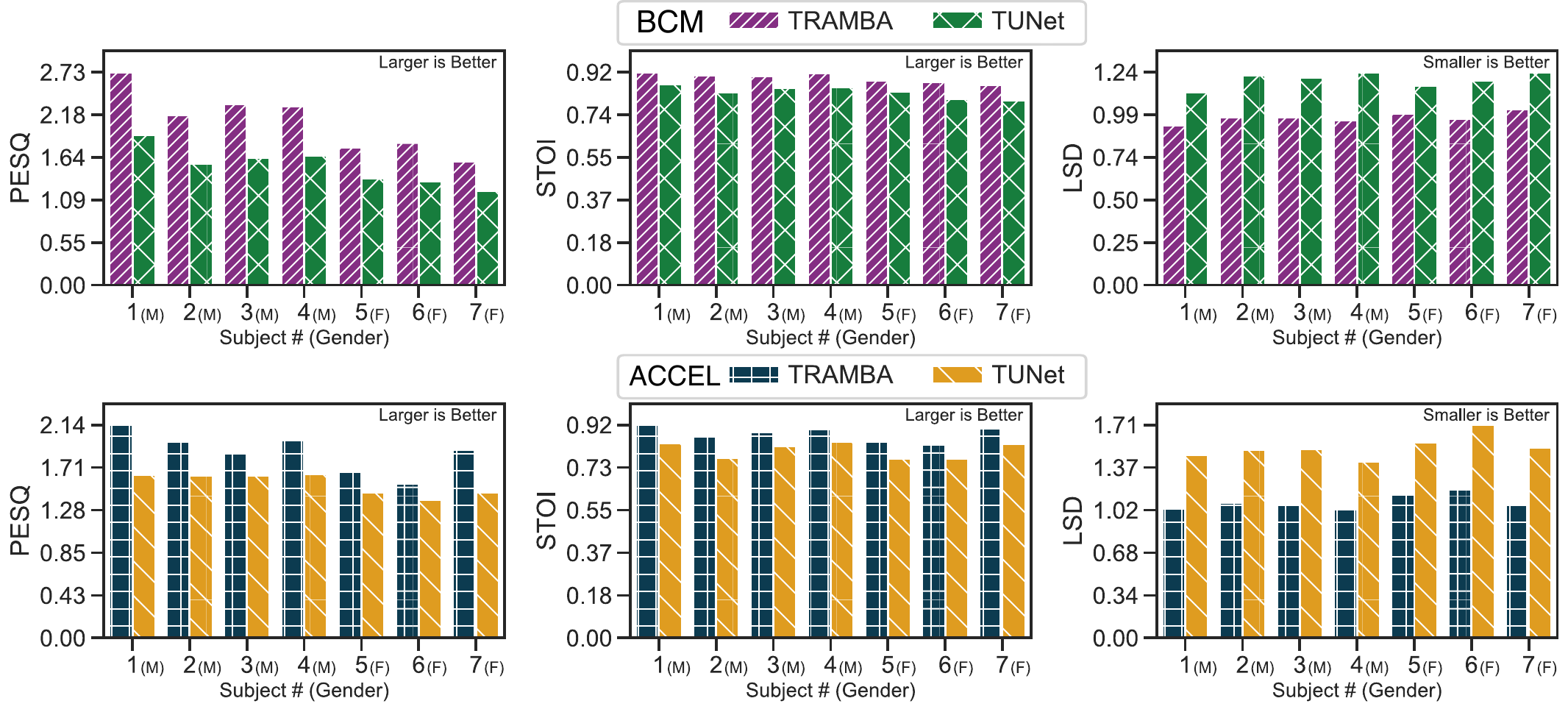}
    \caption{Performance of BCM/ACCEL speech enhancement breakdown by individuals.}
    \label{fig:transfer_bcm_finetuning_persons}
\end{figure}

Figure~\ref{fig:transfer_bcm_finetuning_persons} shows the performance of \name broken down by each volunteer. The BCM and ACCEL was placed at location 2 (Nasal Bone). Again, \name consistently yielded the best performance metrics across all participants. One interesting trend among all methods is that performance is lower among females than males. We suspect that this is due to their higher frequency voices experiencing greater attenuation through bone and skin. In future work, we plan to explore causes an solutions in more depth.


\subsubsection{Data Availability}
\label{subsubsec:data_availability}

\begin{figure}[!t]
    \centering
    \includegraphics[width=\linewidth]{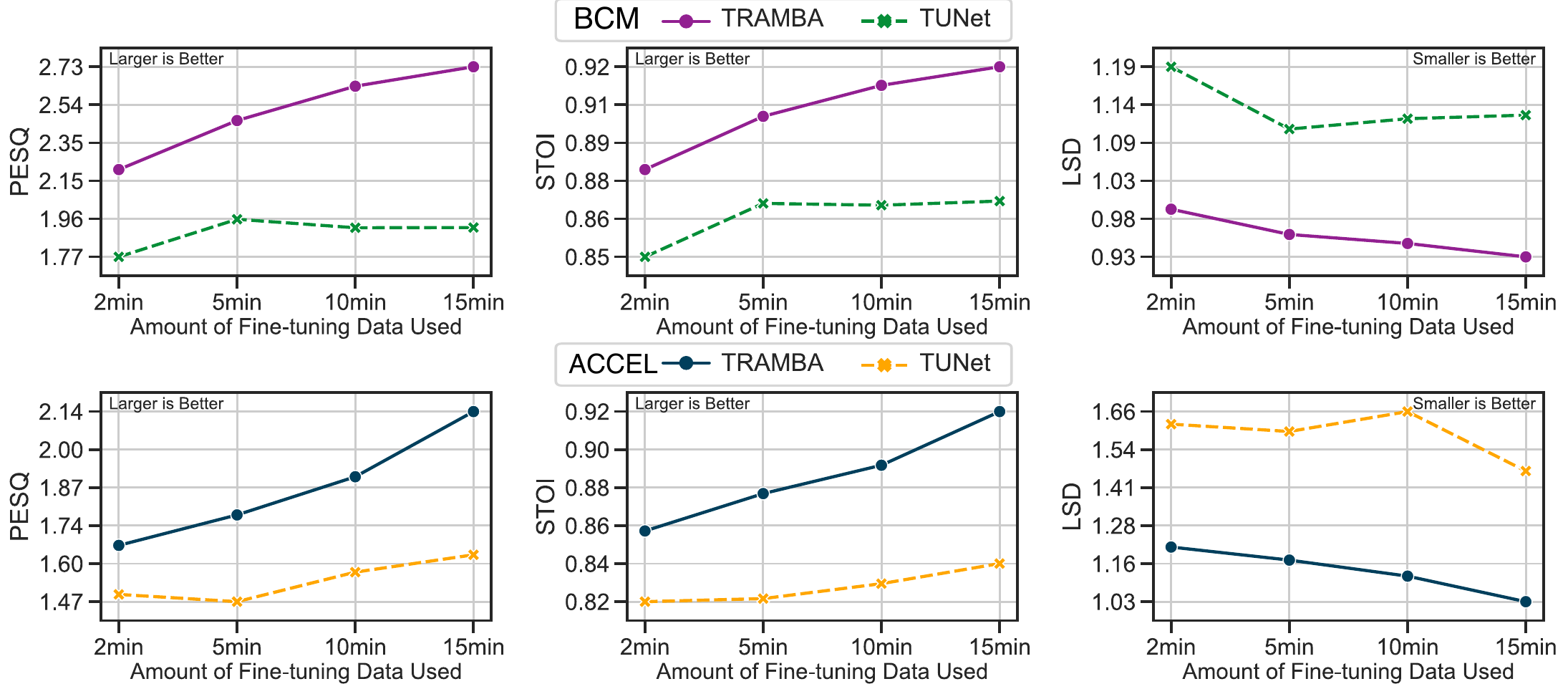}
    \caption{Performance of BCM/ACCEL speech enhancement varying the amount of data available to fine-tune.}
    \label{fig:transfer_bcm_data_amount}
\end{figure}

Figure~\ref{fig:transfer_bcm_data_amount} varies the amount of data we collect from an individual. \name outperforms TUNet with all fine-tuning dataset size. With only $2$ minutes of data, \name outperforms TUNet across all metrics even if TUNet was fine-tuned with $15$ minutes of data collected from the user.


\subsubsection{Fine-tuning Time}
\label{subsubsec:finetuning_time}

\begin{figure}[!t]
    \centering
    \includegraphics[width=\linewidth]{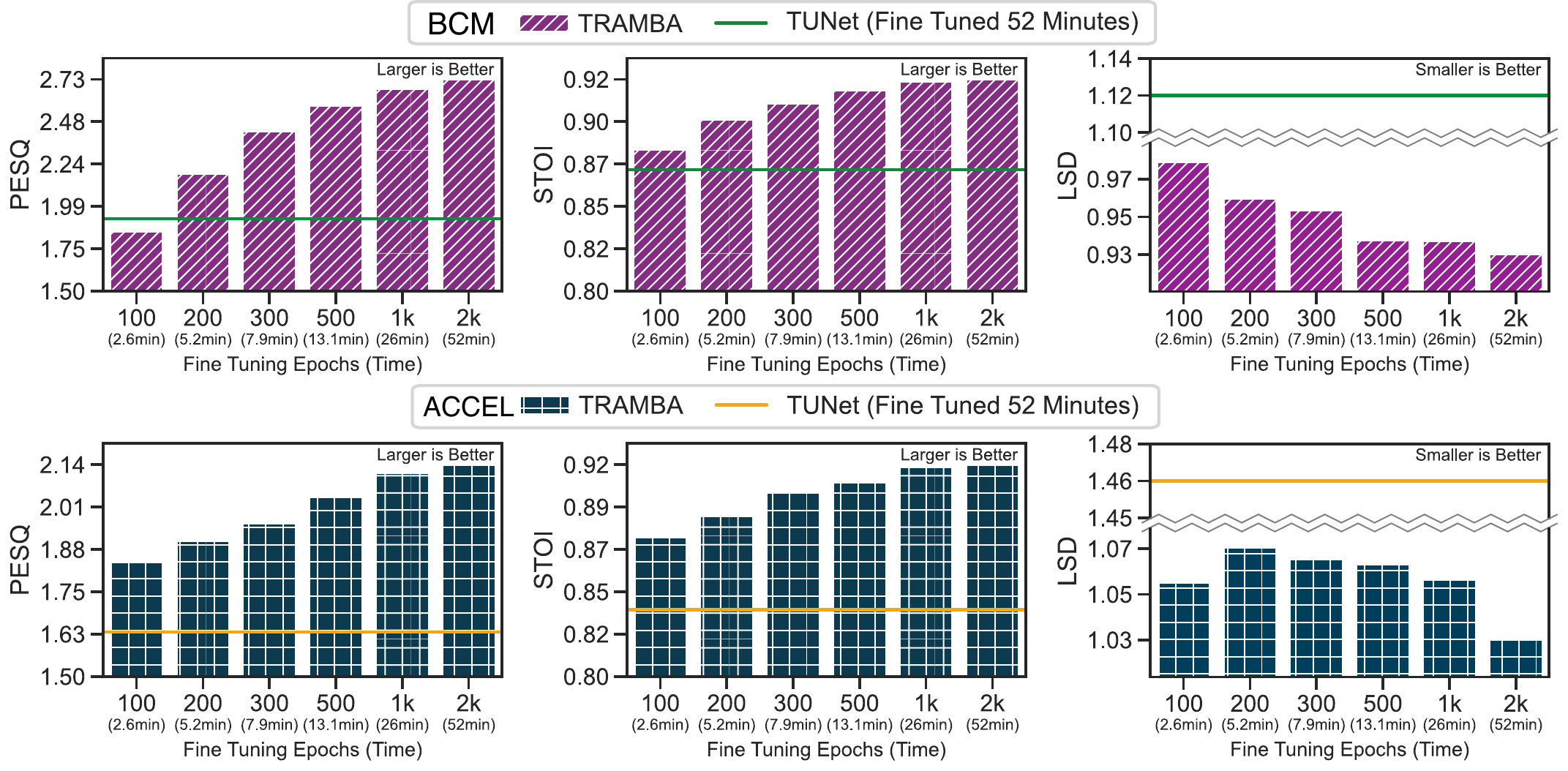}
    \caption{Performance of BCM/ACCEL speech enhancement varying the number of epochs used to fine-tune.}
    \label{fig:transfer_bcm_finetune_time}
\end{figure}

Figure~\ref{fig:transfer_bcm_finetune_time} varies the number of epochs used for fine-tuning up to 2000 epochs, which corresponded to $52$ minutes of training \name. We compare \name to TUNet trained for the same amount of time. Again, \name outperforms TUNet at all training times. With only $2.6$ minutes (100 epochs) of training, \name outperforms TUNet trained for $52$ minutes across all metrics, except for PESQ, which is only slightly lower.


\subsubsection{Sampling Rate}
\label{subsubsec:subsample_rate}

\begin{figure}[!t]
    \centering
    \includegraphics[width=\linewidth]{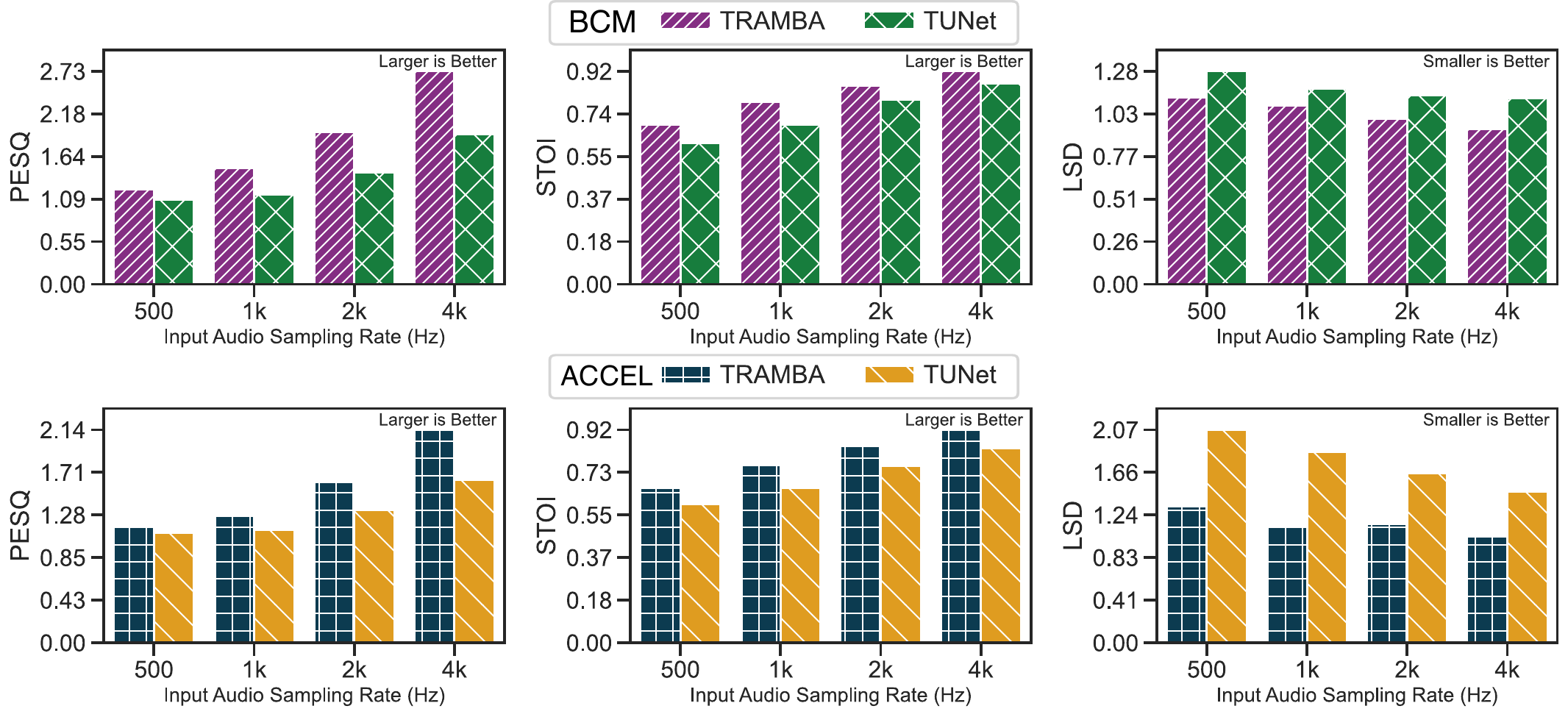}
    \caption{Performance of speech enhancement, varying the the input sampling rate.}
    \label{fig:transfer_bcm_samplerate}
\end{figure}

Figure~\ref{fig:transfer_bcm_samplerate} varies the starting sampling rate to perform super resolution to $16$kHz. \name outperforms TUNet across all starting sampling rates. We observe that \name achieves acceptable speech quality when upsampling from around $2$kHz or higher for both BCM and accelerometer. TUNet only achieves similar performance as \name when the sampling rate is at least double, meaning \name can enhance and generate higher quality and resolution speech with less information, sampling, and transmission bandwidth.


\subsubsection{Unseen Individuals and the Need for Fine-Tuning}
\label{subsubsec:unseen_individuals}

\begin{table*}[!t]
	\centering
	\resizebox{.8\textwidth}{!}{
\begin{tabular}{|c|cccc||cccc|}
\hline
 & \multicolumn{1}{c|}{\textbf{PESQ}} & \multicolumn{1}{c|}{\textbf{STOI}} & \multicolumn{1}{c|}{\textbf{LSD}} & \textbf{WER} & \multicolumn{1}{c|}{\textbf{PESQ}} & \multicolumn{1}{c|}{\textbf{STOI}} & \multicolumn{1}{c|}{\textbf{LSD}} & \textbf{WER} \\ \hline
 & \multicolumn{4}{c||}{BCM - \name} & \multicolumn{4}{c|}{BCM - TUNet} \\ \hline
(i) Pre-train on audio only & \multicolumn{1}{c|}{1.0859} & \multicolumn{1}{c|}{0.6515} & \multicolumn{1}{c|}{1.8639} & 65.51\%& \multicolumn{1}{c|}{1.1141} & \multicolumn{1}{c|}{0.6223} & \multicolumn{1}{c|}{2.2536} & 74.57\%\\ \cline{1-1}
(ii) Creating generalizable model & \multicolumn{1}{c|}{1.1614} & \multicolumn{1}{c|}{0.7544} & \multicolumn{1}{c|}{1.1713} & 34.67\%& \multicolumn{1}{c|}{1.1130} & \multicolumn{1}{c|}{0.7239} & \multicolumn{1}{c|}{1.3392} & 44.52\%\\ \cline{1-1}
(iii) Fine-tuning to the individual & \multicolumn{1}{c|}{2.7273} & \multicolumn{1}{c|}{0.9221} & \multicolumn{1}{c|}{0.9329} & 2.65\%& \multicolumn{1}{c|}{1.9195} & \multicolumn{1}{c|}{0.8701} & \multicolumn{1}{c|}{1.1220} & 6.21\%\\ \hline
 & \multicolumn{4}{c||}{ACC - \name} & \multicolumn{4}{c|}{ACC - TUNet} \\ \hline
(i) Pre-train on audio only & \multicolumn{1}{c|}{1.1893} & \multicolumn{1}{c|}{0.7095} & \multicolumn{1}{c|}{2.1775} & 33.33\%& \multicolumn{1}{c|}{1.1590} & \multicolumn{1}{c|}{0.6714} & \multicolumn{1}{c|}{2.1154} & 18.69\%\\ \cline{1-1}
(ii) Creating generalizable model & \multicolumn{1}{c|}{1.2926} & \multicolumn{1}{c|}{0.7709} & \multicolumn{1}{c|}{1.3226} & 37.50\%& \multicolumn{1}{c|}{1.3073} & \multicolumn{1}{c|}{0.7515} & \multicolumn{1}{c|}{1.7991} & 38.92\%\\ \cline{1-1}
(iii) Fine-tuning to the individual & \multicolumn{1}{c|}{2.1365} & \multicolumn{1}{c|}{0.9195} & \multicolumn{1}{c|}{1.0326} & 7.76\%& \multicolumn{1}{c|}{1.6337} & \multicolumn{1}{c|}{0.8366} & \multicolumn{1}{c|}{1.4649} & 13.29\%\\ \hline
\end{tabular}
	}
 \caption{Performance with fine-tuning and without fine-tuning.}
	\label{tbl:with_without_finetuning}
\end{table*}

\begin{table*}[!t]
	
	\centering
	\resizebox{0.7\textwidth}{!}{
		\begin{tabular}{|c|c|c|c|c|c|c|c|c|}
              \cline{2-9}
              \multicolumn{1}{c}{} & \multicolumn{4}{|c||}{\textbf{BCM}}& \multicolumn{4}{c|}{\textbf{ACCEL}} \\
		   \cline{2-9}
              \hline
			 Method & PESQ & STOI & LSD & \multicolumn{1}{c||}{WER} & PESQ & STOI & LSD & WER \\
             \hline

             TFiLM & 1.0827 & 0.6875 & 1.5167 & \multicolumn{1}{c||}{73.36\%} & 1.3512 & 0.7378 & 1.9940 & 34.31\% \\
             AFiLM & 1.0979 & 0.6903 & 1.5710 & \multicolumn{1}{c||}{72.99\%} & 1.2900 & 0.7433 & 1.9806 & 39.73\% \\
             TUNet & 1.1130 & 0.7239 & 1.3392 & \multicolumn{1}{c||}{44.52\%} & 1.3073 & 0.7515 & 1.7991 &  38.92\%\\
             ATS-UNet & 1.0967 & 0.5717 & 2.1571 & \multicolumn{1}{c||}{81.15\%} & 1.1371 & 0.6583 & 1.7805 & 76.52\% \\
             Aero & 1.1557 & 0.7084 & 1.2218 & \multicolumn{1}{c||}{54.42\%} & 1.1944 & 0.7661 & 1.4720 & 23.45\% \\
             \textbf{\name} & \textbf{1.1614} & \textbf{0.7544} & \textbf{1.1713} & \multicolumn{1}{c||}{\textbf{34.67\%}} & \textbf{1.2926} & \textbf{0.7709} & \textbf{1.3226} & \textbf{37.50\%} \\
             \hline
		\end{tabular}
	}

 \caption{Performance of BCM/ACCEL speech enhancement on unseen individuals after training with $6$ volunteers and testing on an unseen volunteer. The poor performance across all methods suggests that creating a generalizable model with a limited amount of data, due in part to the labor intensive data collection process, is not feasible. Rather, fine-tuning to specific users is currently more practical.}
 \label{tbl:transfer_bcm_finetuning_unseen}
\end{table*}

For completeness, we explore the need for fine-tuning by looking at two cases: (i) directly applying the model pre-trained on audio speech (Section~\ref{sec:superresolution_evaluation}) and (ii) creating a model for BCM and accelerometer speech that is generalizable to unseen individuals.

\noindent
\textbf{(i) Applying audio speech pre-trained model to BCM and ACCEL.} Table~\ref{tbl:with_without_finetuning} shows the performance of the direct application of \name that is only pre-trained on downsampled audio speech (Section~\ref{sec:superresolution_evaluation}) to enhancing BCM and accelerometer speech. PESQ and STOI see very low values, suggesting that training \name using only OTA audio is not sufficient.

\noindent
\textbf{(ii) Creating a generalizable model.} Going one step further, we attempt to create a generalizable model by fine-tuning \name with as big of a dataset as possible. Table~\ref{tbl:transfer_bcm_finetuning_unseen} compares \name with baseline models after training the models with most of our volunteers (6 volunteers) and testing on the unseen (1 volunteer). All methods see equally poor performance, suggesting the need for more data from more individuals to create a general model. Table~\ref{tbl:with_without_finetuning} further highlights the disparity in performance between simply pre-training on audio (i), attempting to create a generalizable model (ii), and fine-tuning to the individual (iii). Until tools and platforms exist for collecting large amounts of paired audio and vibration data, creating and incorporating a general model is not practically feasible.

%% file: sections/system_design.tex
\section{System Design and Implementation with \name}
\label{sec:system_design}

\begin{figure}[!t]
    \centering
    \includegraphics[width=0.60\linewidth]{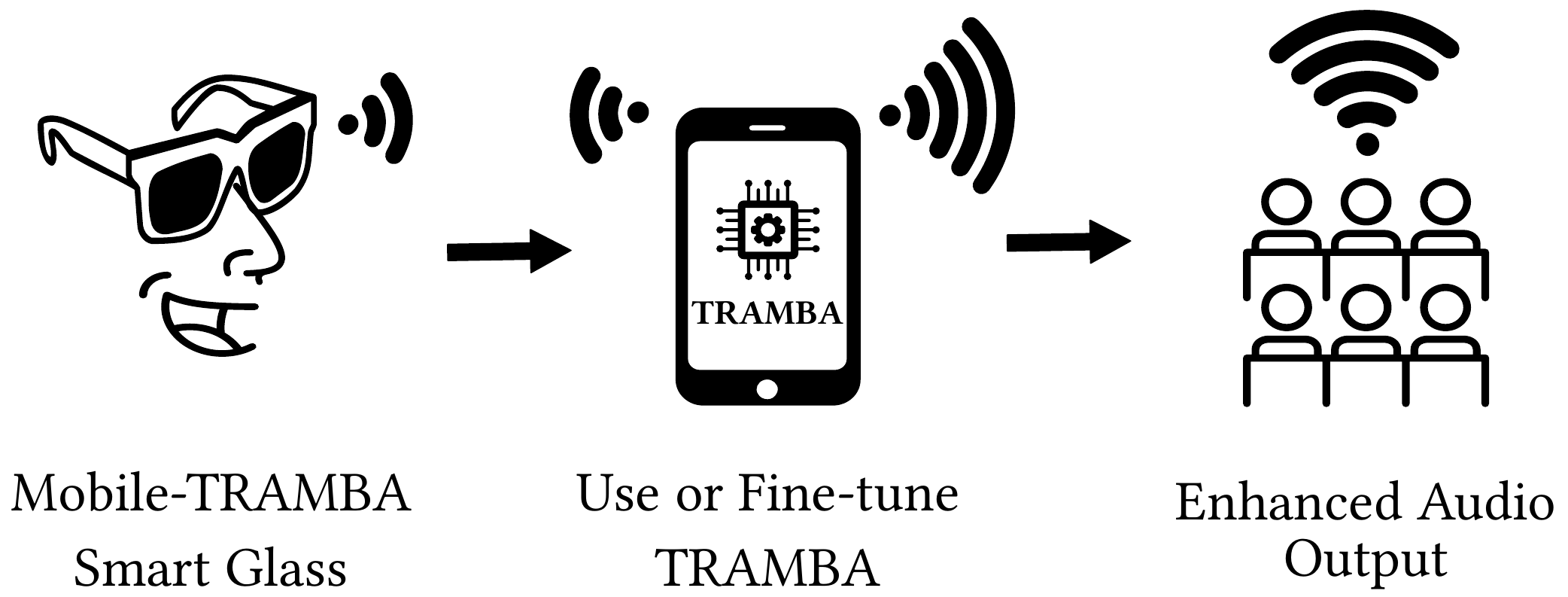}
    \caption{Architecture of the mobile and wearable platform (mobile-\name) that leverages \name to enhance vibration-based speech.}
    \label{fig:system_architecture}
\end{figure}

We incorporate \name into a vibration-based mobile and wearable system, which we call mobile-\name for the rest of the paper, shown in Figure~\ref{fig:system_architecture}. Audio is sampled by the head-worn wearable that is transmitted via Bluetooth Low Energy to the smartphone platform or computer, depending on which device the user is currently using. The smartphone or computer takes incoming data and runs \name. To implement \name on a smartphone, we convert our PyTorch model to a TensorFlow Lite (tflite) model by converting to an ONNX model and a standard TensorFlow model as intermediate steps.

\noindent
\textbf{Wearable Platform:} The wearable prototype we built uses the Seeed Studio XIAO nRF52840 platform, which contains a Nordic nRF52840 Bluetooth Low Energy (BLE) SoC, to transmit speech that is sensed from from an accelerometer or a BCM to the smartphone. For our user studies, we built a case for the wearable such that the accelerometer or BCM is situated on the bridge of the nose, which could be possible if the ACCEL or BCM is placed in the nose pads of a glasses wearable, as demonstrated here~\cite{maruri2018v}. We chose location 2 (Nasal Bone) as shown in Figure \ref{fig:datacollection_setup} for our user studies and end-to-end evaluation because we saw the most consistent performance for both the BCM and ACCEL across all methods at this location. However we believe this platform can be adapted to a multitude of locations, as evidenced in Section~\ref{sec:eval_transfer_learning}.

\noindent
\textbf{Fine-tuning:} The first time a user puts on the wearable, they collect up to 15 minutes of their own voice for fine-tuning. The fine-tuning process collects paired OTA speech and vibration-based speech from the smartphone's microphone and the BCM or accelerometer on the wearable. For the computer version of mobile-\name, the fine-tuning occurs directly onboard. The computer we used has access to an NVIDIA L40 GPU~\cite{l402023nvidia}. Tools for fine-tuning deep learning models on a mobile phone are currently extremely limited. As such, for the smartphone version of mobile-\name, we transmit the paired data to our GPU-equipped computer to fine-tune and update the model. Once the model is fine-tuned, it is transmitted back to the smartphone, where inference occurs locally. In future work, we plan to explore methods for fine-tuning on-device.


%% file: sections/end_to_end_evaluation.tex
\section{End-to-End Evaluation}
\label{sec:end_to_end_evaluation}

Our recruited volunteers (Section~\ref{subsec:eval_tl_data_collection}) fine-tuned the mobile-\name system with 15 minutes of their own voice before reciting 8 minutes of passages from~\cite{BasicSpokenEnglish} across various environments with unique noise profiles at different walking speeds. We compare the performance of \name against over-the-air systems. Namely, we compare against DCUNet~\cite{choi2018phase}, a deep learning model for speech denoising, and Conv-TasNet~\cite{luo2019conv}, a sound source separation deep neural network that extracts speech from OTA audio and the backbone of the ClearBuds earable platform~\cite{chatterjee2022clearbuds} that leverages deep learning to denoise and enhance OTA speech.

The primary quantitative metric we use for comparison is the word error rate (WER) because it is not possible to obtain ground truth clean speech in naturally noisy environments. However, if speech is properly enhanced by significantly attenuating ambient noises and/or sufficiently generating the higher frequency formants of speech, then a speech-to-text model should naturally translate a larger portion of words correctly. All WER metrics were computed after performing super resolution on BCM or ACCEL sampled speech from $4kHz$ to $16kHz$.

\subsection{Performance in Different Environments}
\label{subsec:performance_in_different_environments}
\begin{table*}[!t]
	\centering
	\resizebox{.9\textwidth}{!}{
		\begin{tabular}{|c|c|c|c|} \hline 
              \textbf{Method}& \textbf{Cafeteria (SNR: -0.355 dB)} & \textbf{Loud Music (SNR: -3.742 dB)}& \textbf{Construction Site (SNR: -10.458 dB)} \\ \hline 
		   Raw Audio& 15.85\%& 45.12\%& 86.59\%\\
              Conv-TasNet~\cite{luo2019conv}& 79.88\%& 20.12\%& 85.37\%\\
              DCUNet~\cite{choi2018phase}& 17.07\%& 83.54\%&86.59\%\\
              \name - BCM & 7.93\%& 6.10\%& 14.63\%\\
              \name - ACCEL & 4.27\%& 16.46\%& 10.37\%\\
              \hline
		\end{tabular}
}

 \caption{Word error rate in different environments.}
 \label{tbl:evaluation_environments}
\end{table*}

We recorded 12 total minutes of data across all modalities (BCM, ACCEL, OTA) in various environments shown in Figure~\ref{fig:datacollection_setup}b. The environments we chose had varying types and levels of background noise. Table~\ref{tbl:evaluation_environments} compares the WER between \name, raw audio, and denoising over-the-air signals. There was almost no background noise in the office setting, yielding a WER of almost $0$ across all methods, so we leave out this column. As the noise level increases in the different environments, the WER increases steadily when no processing is performed. Additionally, applying denoising and sound source separation on OTA signals does not always yield a lower WER due to distortions that such models commonly introduce when removing noise~\cite{distortion}. On the other hand, \name sees much higher and consistent performance across all noise settings, even when the noise is 10 orders of magnitude louder. Bone conduction and vibration-based sensing modalities are naturally not sensitive to ambient noises, unlike OTA microphones.


\subsection{Movement}
\label{subsec:movement}

\begin{figure}[!t]
    \centering
    \includegraphics[width=0.90\linewidth]{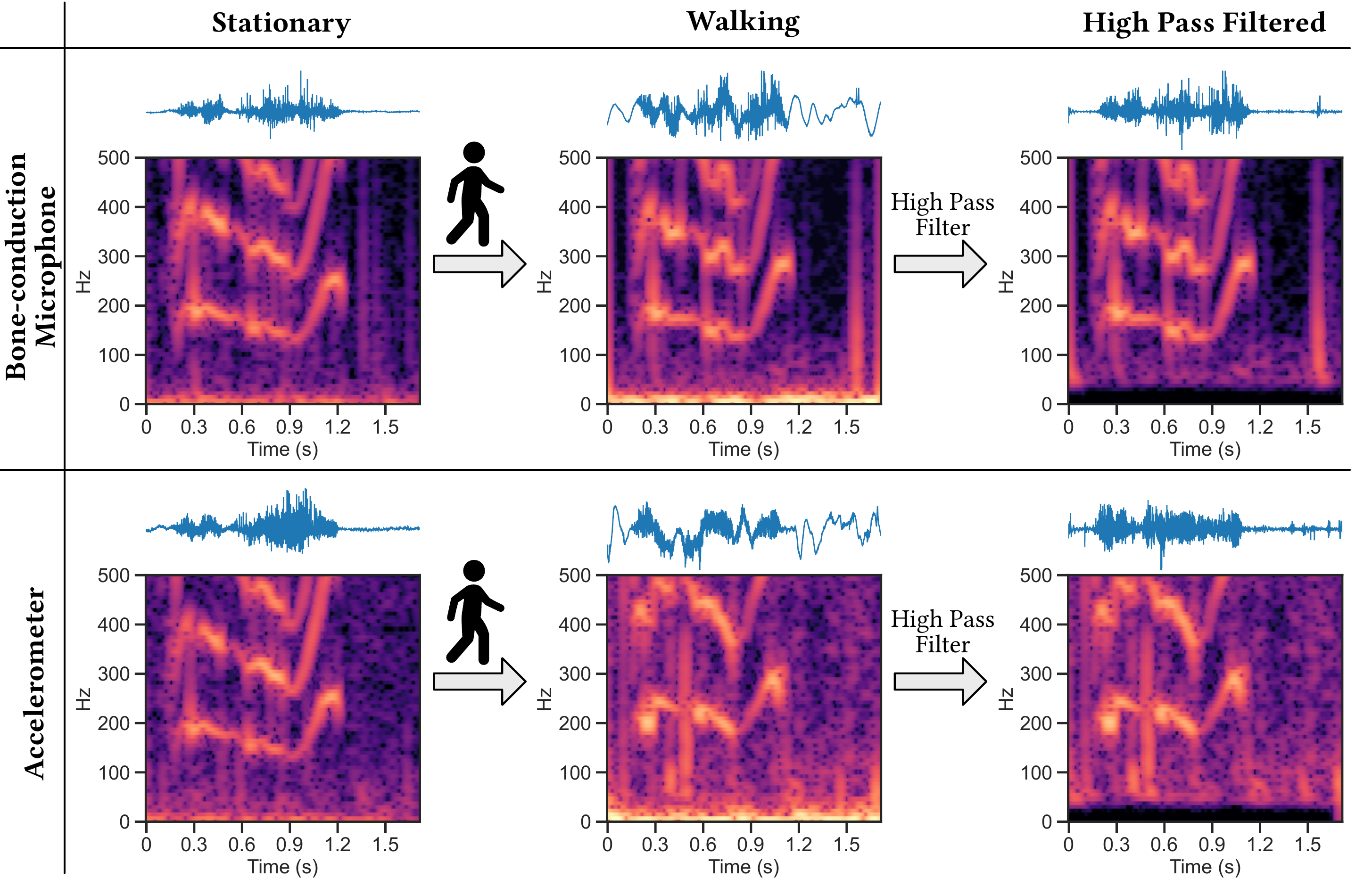}
    \caption{Spectrogram of ACCEL and BCM while user is moving. There is only a small amount of energy in the extreme low frequencies, well below the lowest frequencies of speech (Typically around $100$Hz). This means, movement has little to no effect on performance if a high pass filter is applied (Table~\ref{tbl:movement}).}
    \label{fig:spectrogram_movement}
\end{figure}

\begin{table*}[!t]
    \centering
    \resizebox{.8\textwidth}{!}{
\begin{tabular}{|c|ccc||ccc|}
\hline
 & \multicolumn{1}{c|}{\textbf{PESQ}} & \multicolumn{1}{c|}{\textbf{STOI}} & \textbf{LSD} & \multicolumn{1}{c|}{\textbf{PESQ}} & \multicolumn{1}{c|}{\textbf{STOI}} & \textbf{LSD} \\ \hline
 & \multicolumn{3}{c||}{BCM - \name} & \multicolumn{3}{c|}{BCM - TUNet} \\ \hline
Standing Still & \multicolumn{1}{c|}{1.8063} & \multicolumn{1}{c|}{0.8114} & 1.0814 & \multicolumn{1}{c|}{1.5064} & \multicolumn{1}{c|}{0.7772} & 1.1403 \\ \cline{1-1}
Walking - 1.2 m/s & \multicolumn{1}{c|}{1.7639} & \multicolumn{1}{c|}{0.8031} & 1.0801 & \multicolumn{1}{c|}{1.4956} & \multicolumn{1}{c|}{0.7665} & 1.1511 \\ \hline
 & \multicolumn{3}{c||}{ACC - \name} & \multicolumn{3}{c|}{ACC - \name} \\ \hline
Standing Still & \multicolumn{1}{c|}{1.8300} & \multicolumn{1}{c|}{0.8480} & 1.2461 & \multicolumn{1}{c|}{1.7077} & \multicolumn{1}{c|}{0.8044} & 1.4213 \\ \cline{1-1}
Walking - 1.2 m/s & \multicolumn{1}{c|}{1.7390} & \multicolumn{1}{c|}{0.8413} & 1.2722 & \multicolumn{1}{c|}{1.6888} & \multicolumn{1}{c|}{0.8021} & 1.4323 \\ \hline
\end{tabular}
    }
 \caption{Performance under movement.}
    \label{tbl:movement}
\end{table*}

Figure~\ref{fig:spectrogram_movement} shows the spectrogram of signals collected by the BCM and ACCEL while the user is walking. In this scenario, high energy frequencies appear in the single Hz range or lower, even when the user is running and moving faster, which is significantly lower than the typical lowest frequencies of human voice (approximately $100$ Hz). This suggests that applying a high pass filter would be sufficient to remove the effects of movement from speech, as shown in Table~\ref{tbl:movement}. Here, we applied a first order Butterworth filter with a $10$Hz cut-off frequency. Performance across all metrics remain similar when the user is moving.

\subsection{Power and Latency}
\label{subsec:power_latency}

\begin{table}[ht]
  \centering
  \begin{tabular}{|c|c|c|}
    \hline
     & Data Rate& Power Consumption\\ \hline
    \name - 500 Hz& 8 kbps& 2.49 mW\\
    \name - 1 kHz& 16 kbps& 2.58 mW\\
    \name - 2 kHz& 32 kbps& 2.75 mW\\
    \name - 4 kHz& 64 kbps& 3.21 mW\\
    \name - 8 kHz& 128 kbps& 4.09 mW\\
    No processing (16 kHz)& 256 kbps& 6.48 mW\\ \hline
  \end{tabular}
\caption{Total power consumption of \name's wearable, while sampling and streaming BCM/ACCEL data at different frequencies to the smartphone. Sampling at $4$kHz more than halves the power consumption from sampling and transmitting at full resolution ($16$kHz).}
\label{tab:power_earbuds}
\end{table}

\begin{table}
    \centering
    \begin{tabular}{|c|c|c|c|c|} \hline 
         Device&  iPhone 15 Pro&  iPhone 14 Pro&  iPhone 13 Pro&iPhone 12\\ \hline 
         Inference Time (ms)&  19.584&  20.343&  22.992&27.931\\ \hline
    \end{tabular}
    \caption{Inference time (latency) on various phone models. \name processes windows of $512$ms at a time and can perform inference in real-time on most modern smartphones.}
    \label{tbl:phone_latency_power}
\end{table}
             
              
             
              
             
              

Table~\ref{tab:power_earbuds} shows the power consumption of sampling and transmitting data from the wearable to the smartphone platform at different sampling rates. Performing speech enhancement with \name on $4kHz$ signals us to sample and transmit $75\%$ less data than sampling high resolution, $16kHz$, OTA speech and reduce power consumption by more than $50\%$. We achieve even more aggressive savings if the resolution is further decreased.

Table~\ref{tbl:phone_latency_power} shows the average inference time for one window ($512$ms) on various smartphones. On all tested devices, the inference time is less than $30$ms. Since we process $512$ms windows, \name can perform speech enhancement on many modern smartphones in real-time.

\subsection{User Study}
\label{subsec:user_study}

\begin{figure}[!t]
    \centering
    \includegraphics[width=0.65\linewidth]{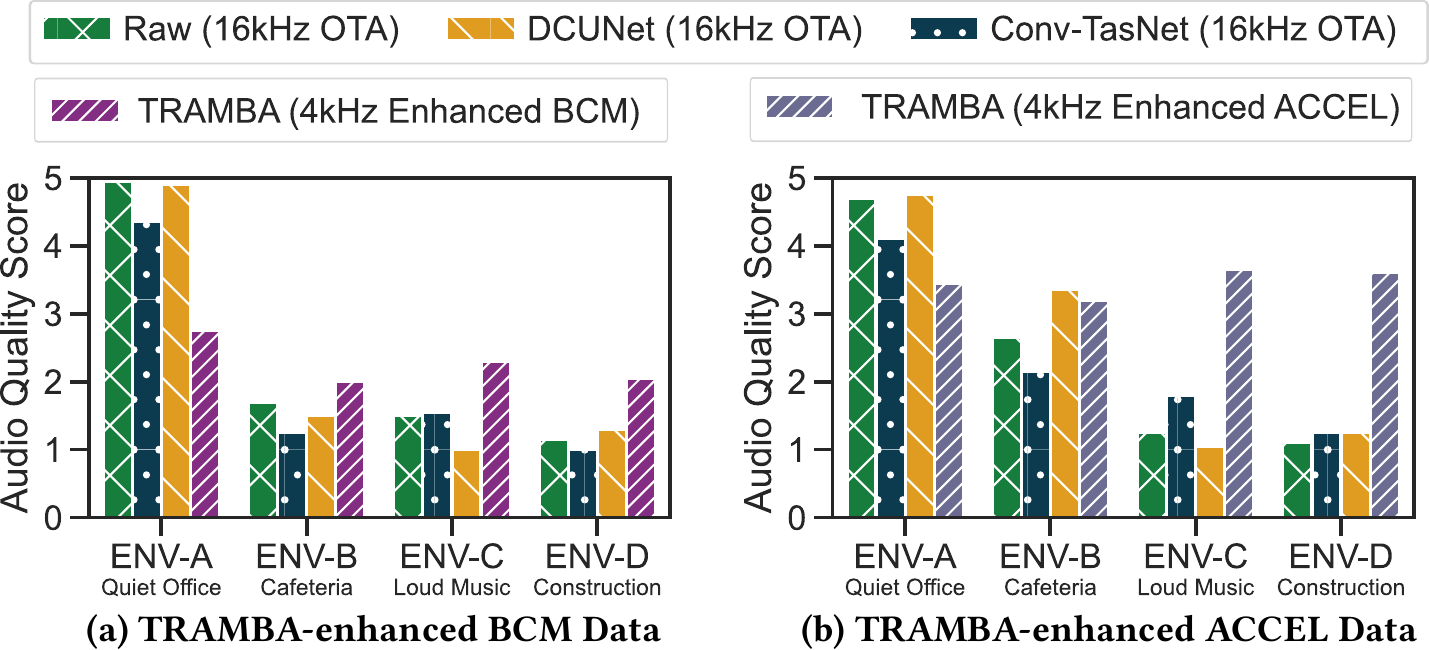}
    \caption{Perceived quality evaluated by users. Higher number denotes better audio quality.}
    \label{fig:user_study}
\end{figure}

To evaluate the sound quality in the noisy environments, we asked our recruited volunteers to rate the enhanced audio on a 5 point Likert scale~\cite{joshi2015likert}, where a score of 1 is very poor and 5 is very high quality. Figure~\ref{fig:user_study} shows these results across all environments compared to raw audio and OTA source separation methods. The over-the-air methods were perceived to have higher quality in the office environment because were no loud ambient noises interfering with speech. However, environments B, C, D contained increasingly louder ambient backgrounds. We see that the perceived quality of \name remained fairly constant in all scenarios, demonstrating its lack of sensitivity to ambient noises. However, the quality of speech generated by over-the-air methods drastically decreases, as ambient noise overpowers OTA speech.


%% file: sections/discussion.tex
\section{Discussion}
\label{sec:discussion}

We have demonstrated \name, a novel hybrid transformer and Mamba-based architecture for speech super resolution that outperforms existing state-of-art speech enhancement methods across multiple acoustic modalities (over-the-air, bone-conduction microphones, and accelerometers). Through integrating and deploying \name into real systems and environments, we show that \name enables practical deployment of acoustic in mobile and wearable platforms by surpassing the performance gap with a memory footprint that is orders of magnitude less than the state-of-art, reducing reliance on paired data that is labor-intensive to collect by pre-training on easily obtainable downsampled clean speech audio, and leveraging the reduced data rate required to lower power consumption and improve battery life. While our results show great promise, there are several lines of future work that we believe will further strengthen the practical adoption of acoustic speech enhancement in mobile and wearable platforms.

\noindent
\textbf{On-Device Fine-tuning.} In this work, we did not fine-tune \name directly on the smartphone due to a lack of frameworks that allow developers to fine-tune directly on the phone. We plan to explore efficient on-device fine-tuning and create open-source, easy-to-use tools to enable developers, engineers, and the greater community to fine-tune large deep learning models on smartphone and other edge platforms. 

\noindent
\textbf{Generalizability.} While fine-tuning the pre-trained model with the acoustic modality of choice at box opening allows us to adapt \name to a specific user, generalizing the model to unseen individuals without fine-tuning is still unsolved. We plan to build upon the open-source platforms and techniques created in this work to further reduce the barrier for collecting paired vibration and OTA data. In doing so, we hope to enable community-driven activity that will increase the amount of publicly available data needed to train and create generalizable vibration-based speech enhancement models. Moreover, we observed a degradation in performance in females compared to males, which we hypothesize is due to their naturally higher pitched voices experiencing more severe attenuation through skin and bone. In future work, we plan to explore causes and solutions in more depth.

\noindent
\textbf{Security and Privacy.} In this work, we transmit sensitive speech signals between multiple devices. In future work, we plan to explore privacy-aware transmission schemes that would enable safe and reliable communication of sensitive speech between multiple devices. Moreover, we plan to explore other methods of data reduction, beyond sampling at a lower rate (e.g., compression and learning latent embeddings), that would maintain or further reduce power consumption while improving privacy guarantees.

%% file: sections/conclusion.tex
\section{Conclusion}
\label{sec:conclusion}

We present \name, a novel hybrid transformer and Mamba-based architecture for speech super resolution and enhancement for mobile and wearable platforms. One of the biggest challenges in creating speech enhancement methods for vibration-based sensing modalities is the lack of publicly available datasets arising from the labor-intensive data collection process required. To overcome this challenge, we pre-train \name on widely available, downsampled, clean speech audio before fine-tuning on a small amount of data collected from the user upon a one-time initial setup. We demonstrate that \name outperforms existing state-of-art speech super resolution methods across multiple acoustic modalities (over-the-air, BCM, and accelerometer) with a memory footprint of only $19.7$ MBs, compared to GANs that require at least hundreds of MBs or more. We integrate \name into a real mobile and head-worn wearable system and show, through real deployments and user studies, that \name can generate higher quality speech, in noisy environments, compared to systems that denoise audio collected by microphones that often hear ambient background noises. On the systems side we demonstrate that \name can improve the battery life of wearable systems, by up to 160\%, by reducing the resolution of audio that needs to be sampled and transmitted. \name is a critical step towards practical integration of speech enhancement into mobile and wearable platforms.